\documentclass[traditabstract]{aa}                                    \usepackage{amsmath}

\usepackage[varg]{txfonts}
\usepackage{graphicx}
\usepackage{natbib}
\usepackage{url}
\usepackage{flafter}
\usepackage{longtable}

\bibpunct{(}{)}{;}{a}{}{,}
\begin{document}
   \title{The rate of supernovae at redshift 0.1-1.0}

   \subtitle{-- the Stockholm VIMOS Supernova Survey \rm{\bf{IV}} 
   \thanks{Based on observations collected at the European Organisation
   for Astronomical Research in the Southern Hemisphere, Chile, under ESO 
   programme ID 167.D-0492.}}
   \author{J. Melinder \inst{1} \and
        T. Dahlen \inst{2} \and
	L. Menc\'ia Trinchant \inst{1} \and
	G. \"Ostlin \inst{1} \and
	S. Mattila \inst{1,3} \and
        J. Sollerman \inst{1} \and
	C. Fransson \inst{1} \and
      	M. Hayes \inst{4} \and
	E. Kankare \inst{3} \and 
        S. Nasoudi-Shoar \inst{5}
        }
    \institute{Department of Astronomy, Oskar Klein Centre, Stockholm University, AlbaNova 
               University Centre, SE-106 91 Stockholm, Sweden\\ 
               \email{jens@astro.su.se}
		 \and
		 	 Space Telescope Science Institute, 3700 San Martin Drive, 
             Baltimore, MD 21218, USA
         \and
             Tuorla Observatory, Department of Physics and Astronomy, University of Turku, 
             V\"ais\"al\"antie 20, FI-21500 Piikki\"o, Finland
         \and
			 CNRS; Universit\'e de Toulouse, UPS-OMP, IRAP, Toulouse, France
         \and
             Argelander-Institut f\"ur Astronomie, Universit\"at Bonn, 
             Auf dem H\"ugel 71, 53121 Bonn, Germany}
   \date{Received ; accepted }
  \abstract{
We present supernova rate measurements at redshift 0.1--1.0 from the Stockholm
VIMOS Supernova Survey (SVISS). The sample contains 16 supernovae in total. The
discovered supernovae have been classified as core collapse or type Ia supernovae (9 and
7, respectively) based on their light curves, colour evolution and host galaxy
photometric redshift. The rates we find for the core collapse supernovae are
$3.29_{-1.78\,-1.45}^{+3.08\,+1.98}\times 10^{-4}$ yr$^{-1}$ Mpc$^{-3}$ h$^3_{70}$ (with
statistical and systematic errors respectively) at average redshift 0.39 and 
$6.40_{-3.12\,-2.11}^{+5.30\,+3.65}\times 10^{-4}$ yr$^{-1}$ Mpc$^{-3}$ h$^3_{70}$ at
average redshift 0.73. For the type Ia supernovae we find a rate of
1.29$_{-0.57\,-0.28}^{+0.88\,+0.27}\times 10^{-4}$ yr$^{-1}$ Mpc$^{-3}$ h$^3_{70}$ at
$\langle z\rangle =0.62$. All of these rate estimates have been corrected for
host galaxy extinction, using a method that includes supernovae missed in
infrared bright galaxies at high redshift. We use Monte Carlo simulations to
make a thorough study of the systematic effects from assumptions made when
calculating the rates and find that the most important errors come from
misclassification, the assumed mix of faint and bright supernova types and
uncertainties in the extinction correction.  We compare our rates to other
observations and to the predicted rates for core collapse and type Ia supernovae
based on the star formation history and different models of the delay time
distribution.  Overall, our measurements, when taking the effects of extinction
into account, agree quite well with the predictions and earlier results. 
Our results highlight the importance of understanding the role of systematic
effects, and dust extinction in particular, when trying to estimate the rates
of supernovae at moderate to high redshift.}
   \maketitle
\section{Introduction}
The cosmic rate of supernovae is an important observable that can be
used to constrain the properties of galaxies at high redshifts and to
study the supernovae (SNe) themselves.  Depending on the origin of the
SN explosion, thermonuclear or core collapse, different aspects
can be studied. The first measurement of the cosmic SN rate (SNR) was
done by \citet{1938ApJ....88..529Z} who found that ``the average
frequency of occurrence of supernovae is about one supernova per
extra-galactic nebula per six hundred years'' for the local volume. It is
not until the latest decades that the higher redshift regimes have been
possible to reach.

More recently, large programmes targeting type Ia supernovae at intermediate and
high ($\gtrsim 0.1$) redshifts have been conducted to measure the expansion of
the universe and do precision cosmology \citep[e.g.
][]{1997ApJ...483..565P,1998ApJ...507...46S,
2006A&A...447...31A,2007ApJ...659...98R,2007ApJ...666..674M}. Some of these
surveys also report supernova rates for Ia SNe
\citep[e.g.][]{2006AJ....132.1126N,2008ApJ...681..462D} out to $z\sim 1.5$ and
core collapse supernovae (CC SNe)
\citep{2004ApJ...613..189D,2009A&A...499..653B} out to $z\sim 0.7$.  Large
surveys targeting SNe of any type have also been successful in finding and
characterising supernovae as well as determining both local and intermediate
redshift cosmic supernova rates \citep{2008A&A...479...49B,2010ApJ...713.1026D,
2011MNRAS.tmp..413L}.  \citet{2010ApJ...718..876S},
\citet{2010ApJ...715.1021D}, and \citet{2012ApJ...745...32B} survey galaxy
clusters, where the SN Ia rates are likely to be enhanced, to find supernovae
and have reported cluster SN rates out to redshift 0.9.
\citet{2012ApJ...745...31B} also report SN Ia rates from detections in the
foreground and background of the targeted galaxy clusters out to $z\sim 1.5$.

The standard observational strategy for finding SNe at high redshift is
to perform survey observations on a given field and then monitor the same field
over a long period of time. Supernovae are discovered by searching the images
for variable sources using image subtraction tools \citep[such as
][]{2000A&AS..144..363A} to minimise subtraction residuals.  The cadence of
observations during the survey period is normally chosen to sample signature
features of the SN light curves and colour evolution at the target redshifts.
In this way photometric typing of the SNe is possible and the light curves can
be used to study the SN characteristics. For the Ia surveys with cosmology as
the main goal, follow-up spectroscopy of SN Ia candidates is needed to get a
secure determination of the redshift, to improve the accuracy in the distance
measurement, and confirmation of the type. When calculating supernova rates
from this kind of survey, care has to be taken to avoid selection effects from
the spectroscopic observations.

Supernova typing normally includes studying the spectra of the SNe close to
their peak luminosity and identifying spectral lines, notably H, He and
Si\rm{II} lines, something which is observationally very expensive and in
practice unfeasible at high redshift for fainter SN types. Another method to
type SNe is to compare the observed light curves and colour evolution to
pre-existing templates of different SN types, i.e. photometric SN typing. These
methods have been demonstrated to work \citep[e.g.
][]{2007ApJ...659..530K,2007AJ....134.1285P,2009ApJ...707.1064R,
2010PASP..122.1415K} using somewhat different techniques. Better typing
accuracy is achieved with prior information on the redshift through photometric
or spectroscopic redshift of the host galaxy
\citep[][]{2010PASP..122.1415K,2011A&A...532A..29M}.

Thermonuclear supernovae (or SN Ia's) are thought to be  white dwarfs that
explode when they accrete matter and approach the Chandrasekhar limit
\citep[for a review, see][]{2000A&ARv..10..179L}. When taking the
luminosity-stretch relation \citep{1993ApJ...413L.105P} into account the peak
luminosity of these SNe exhibit a very narrow spread and can thus be used to
accurately measure cosmological distances. The exact details of the explosions
and of the progenitor system are not fully understood. For example, the time
between formation of the progenitor system and the supernova explosion -- the
so called delay time -- is unknown. This delay time depends on the nature of
the companion star to the white dwarf \citep{2005A&A...441.1055G}. By studying
the rates of Ia supernovae and comparing to either the cosmic star formation
history \citep[e.g.][]{1999A&A...350..349D,
2006AJ....132.1126N,2010ApJ...713...32S}, or the star formation rates and
stellar masses of the host galaxies
\citep{2006ApJ...648..868S,2008PASJ...60.1327T,2011MNRAS.tmp..307M} it is
possible to set constraints on the delay time and thereby also on the
progenitor system.

Core collapse supernova explosions (CC SNe) are the end-points of the lives of
massive stars, with masses between $\sim 8 \,\mbox{M}_{\sun}$ and $\sim 50 \,
\mbox{M}_{\sun}$ \citep{1984ApJ...277..791N,
1997ApJ...483..228T,2009ARA&A..47...63S}. Since massive stars are short-lived
compared to the cosmic time-scales the CC SNe trace active star formation.  By
averaging the CC SN rate over cosmic volume the rest frame rate of star
formation in that volume can be studied.  In this way an independent measure of
the star formation history at high redshift can be obtained
\citep{2004ApJ...613..189D,2005A&A...430...83C, 2008A&A...479...49B,
2009A&A...499..653B}. More conventional methods of finding the cosmic star
formation rates include measuring the rest-frame UV light from galaxies at a
given redshift \citep[e.g.  ][]{2004ApJ...600L.103G,2009ApJ...705..936B},
measuring the far-infrared (FIR) light to take star formation hidden by dust
into account \citep[e.g.][]{2005ApJ...632..169L} and deriving the rates from
H$\alpha$ measurements \citep{2009ApJ...696..785S,2010A&A...509L...5H}.  The UV
and H$\alpha$-based methods have a drawback, that is also present for the SNR
method, in that a correction for dust extinction needs to be applied.  Methods
based on using the FIR light to estimate the added star formation from
re-radiated UV light make it possible to correct the star formation history for
dust extinction effects.  \citet{2006ApJ...651..142H} presented a compilation
of the star formation history from multiple sources for $z\sim 0-6$, taking the
obscured star formation into account, and fitted an analytical function to the
data.

The light from a supernova has to pass through its host galaxy before starting
on the long trip to reach our telescopes. When travelling through the gas and
dust inside its host the supernova light will be subject to varying degrees of
extinction, depending on the dust content of the galaxy and the position of the
SN with respect to the observer (e.g. a SN situated in an edge-on galaxy will
suffer from higher extinction, on average, than one in a face-on galaxy).
\citet{1998ApJ...502..177H} and \citet{2005MNRAS.362..671R} present models of
extinction for core collapse and thermonuclear SNe in normal spiral galaxies.
These models are built by using Monte Carlo simulations of supernova positions
within a galaxy with given morphology and dust content. By using the extinction
models it is possible to estimate the effect on the observed supernova rates
\citep{2004ApJ...613..189D}. It should be noted that this method is mainly
applicable to normal galaxies with low to medium amounts of gas/dust, typical
of galaxies in the local volume of the universe.

As the redshift increases, more and more of the star formation takes place
inside dusty galaxies.  \citet{2005ApJ...632..169L, 2009A&A...496...57M,
2011A&A...528A..35M} find that the star formation from these infrared bright
galaxies dominate the global star formation at redshift 0.7 and higher. In
these galaxies the SN explosions can be completely obscured by the large
amounts of dust within the nuclear regions. For low to moderate amounts of
extinction this can be estimated and taken into account in the light curve
analysis. But for host galaxies with high dust content (such as luminous and
ultra-luminous infrared galaxies, LIRGs and ULIRGs) most of the SNe may not be
detectable, even in the near-infrared (NIR) where the extinction is strongly
reduced \citep[e.g.][]{2001MNRAS.324..325M}. When calculating the rates, the
number of missing SNe due to the change in average extinction in star forming
galaxies with redshift needs to be compensated for \citep{2007MNRAS.377.1229M}.
The derivation of the de-bias factors that are needed to correct the rates for
this effect is further complicated by the recent discovery that the population
of U/LIRGs at low redshift is quite different from the ones at high redshift
\citep[e.g.][]{2010ApJ...713..686D, 2011A&A...533A.119E, 2011arXiv1110.4057K}.
\citet{2012arXiv1206.1314M} have recently estimated the fraction of SNe
missed in such galaxies both locally and as a function of redshift making use
of the most recent results on the nature of U/LIRGs at different redshifts.

The Stockholm VIMOS Supernova Survey (SVISS) is a multi-band ($R+I$) imaging
survey aimed at detecting supernovae at redshift $\sim$0.5 and derive
thermonuclear and core collapse supernova rates. The supernova survey data were
obtained over a six month period with VIMOS/VLT.
\citet{2008A&A...490..419M} describe the supernova search method along with
extensive testing of the image subtraction, supernova detection and photometry.
The discovery and typing of 16 supernovae in one of the search fields is
reported in \citet{2011A&A...532A..29M}. Here we present the supernova
rates estimated from the survey data along with delay time distributions for
the Ia SNe and star formation rates (SFR) calculated from the CC SNe rates.

The first part of the paper contains a description of the observations and
supernova sample.  In Sect.~\ref{sec:method} we describe the method used to
determine the supernova rates and in Sect.~\ref{sec:results} the resulting
supernova rates, delay time distribution and star formation rates are
presented. In the final section of the paper, Sect.~\ref{sec:disc}, we
discuss the results and compare them to the work of other authors. The Vega
magnitude system and a standard $\Lambda$CDM cosmology with
$\{H_0,\Omega_{M},\Omega_{\Lambda}\}=\{70,0.3,0.7\}$ have been used throughout
the paper.

\section{The data}
\label{sec:data}
\subsection{Observations}
The data used in this paper were obtained with the VIMOS instrument
\citep{2003SPIE.4841.1670L} mounted on the ESO Very Large Telescope (UT3) at
several epochs during 2003--2006. The VIMOS instrument has four CCDs, each
2k$\times$2.4k pixels with a pixel scale of 0.205\arcsec/pxl, covering a total
area of roughly $4\times$56 sq. arcmin. The observations were obtained in the
ELAIS-S1 field \citep{2004AJ....127.3075L}, in five broad band filters ($U$,
$B$, $V$, $R$ and $I$) centred at $\alpha = 00^h 32^m 13^s$, $\delta = -44\degr
36\arcmin 27\arcsec$ (J2000). The supernova search filters were $R$ and $I$.
Observations in these filters were divided into seven search epochs and 2
additional reference epochs (one before and one after). The search epochs were
separated by roughly one month. The observational programme did not include 
any spectroscopy of the detected SNe and the analysis presented here is based on 
the $R+I$ photometric data exclusively.

The individual frames in each epoch were reduced, including removal of fringes,
registered to a common frame of reference and stacked. Each epoch image was
photometrically calibrated using photometric standard stars observed during one
of the nights and local standard stars selected in the field.  Detailed
measurements of the seeing in each of the frames were done by modelling the
point spread function. For more details on the observations and data reduction,
see \citet{laia} and
\citet{2011A&A...532A..29M}.

\subsection{The supernova sample}
\label{sec:sample}
The supernovae were detected by using a dedicated pipeline that was developed
and thoroughly tested by our team. For a detailed description along with the
results of the testing see \citet{2008A&A...490..419M}.  The pipeline includes
the following steps, first the reference image and the search image are scaled
to a common seeing using the ISIS 2.2 code \citep{2000A&AS..144..363A} and the
reference image subtracted from the search image. Automatic source detection
using SExtractor \citep{1996A&AS..117..393B} together with a by-eye inspection
is then done on the subtracted image to find variable objects. The final step
is the aperture photometry, and point spread function (PSF) modelling based
photometry, of the sources. Detailed simulations of the photometric
accuracy are done to make sure that the error estimates are valid. 

To avoid including spurious detections in the output catalogues we required
that the supernova candidates were detected in both bands in two consecutive
epochs. A detection is here defined as being brighter than the $3\sigma$
limiting magnitude in that epoch.  Furthermore, we used the late control epoch
to remove non-SNe (most likely active galactic nuclei, AGN) from the sample,
since no SNe are expected to be visible $\sim 1$ year after explosion at the
wavelengths and redshifts considered here.  It should be noted that the
decision to require detection in both the $R$ and $I$ filters has a significant
impact on the number of detected SN Ia's. These SNe are inherently redder than
the CC SNe. At $z\gtrsim0.8$ their light starts to become redshifted
beyond the $R$ filter and they are less likely to be detected in both of the
filters used in this survey. This effect is taken into account in the rate
calculations (see Sect.~\ref{sec:method}).  

\begin{table}
\caption{Supernovae in the SVISS}	     \label{table:sne} \centering \begin{tabular}{l l l l l} 
\hline\hline SVISS ID & Type & P(Type) & $z$ \\ \hline 
SVISS-SN43  & Ia & 0.886 &0.43\\ 
SVISS-SN161 & Ia & 1.000  &0.50\\ 
SVISS-SN115 & Ia & 1.000  &0.40\\ 
SVISS-SN116 & Ia & 1.000  &0.55\\ 
SVISS-SN309 & Ia & 1.000  &0.47\\
SVISS-SN402 & Ia & 1.000  &0.22\\ 
SVISS-SN135 & Ia & 0.950 &0.98\\
\hline					 SVISS-SN14 & CC & 0.623& 0.36 \\
SVISS-SN51 & CC & 1.000 & 0.51 \\
SVISS-SN54 & CC & 0.812& 0.77 \\
SVISS-SN261& CC & 0.734& 0.37\\
SVISS-SN55 & CC & 0.995& 0.83 \\
SVISS-SN31 & CC & 0.999& 0.12 \\
SVISS-SN56 & CC & 0.930& 0.57 \\
SVISS-SN357& CC & 1.000& 1.40  \\
SVISS-SN24 & CC & 0.643& 0.81 \\ 
\hline					 
\end{tabular} 
\tablefoot{
P(Type) is the probability (Bayesian evidence) for the best fitting 
main type (TN/CC).}
\end{table}

For the sample of supernovae considered in this paper we used the pipeline
described above on the ELAIS-S1 $R+I$ data and found a total of 16 supernovae,
seven Ia SNe and nine CC SNe, ranging from
$z\sim 0.1$ to $z\sim1.4$ (see Table~\ref{table:sne}). The light curves and
colour evolution of the SNe were used to classify them into either Ia-like or
CC types (9 different subtypes, see Table~\ref{table:templ}). We
then co-added the likelihoods for all subtypes belonging to the type Ia or CC
SNe, respectively, to find the most likely type.  This was done using a
Bayesian model selection code, where priors were used for the redshift (host
galaxy photometric redshift), absolute magnitude distribution, extinction and
time after explosion. In \citet{2011A&A...532A..29M} we describe in detail how
the sample was obtained together with the typing code. In that paper we also
investigated the misclassification errors when applying our code to the SN
sample by using a Monte Carlo simulated sample of $\sim 18000$ SNe and a local
SN sample from SDSS-II containing 87 SNe. The resulting errors are used to
estimate systematic errors for the SN rates (see Sect.~\ref{sec:systerrs}.

Over the full redshift range where we have found SNe ($z\lesssim 1.4$) we 
expect approximately 8\% of the found Ia SNe, and 2\% of the CC SNe, to 
be false positives. On the other hand, it then follows that only about 
2\% of the Ia SNe are mistyped compared to about 8\% of the CC SNe. 
This systematic effect is thus not symmetric, overall we expect the 
Ia count to be slightly inflated and the CC count slightly deflated.
  
\subsection{Supernova redshifts}
\begin{figure}
\centering
\includegraphics[width=8cm]{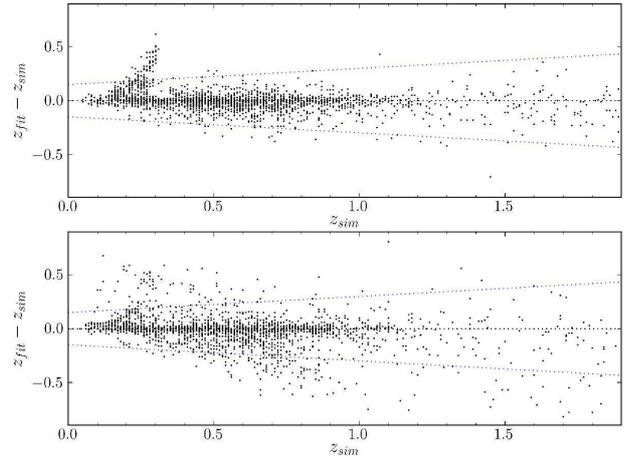}
\caption{The redshift accuracy for simulated supernovae
with (upper panel) and without (lower panel) a host galaxy photometric
redshift. The $\sim 2000$ SNe of all different subtypes used to make
this figure are a sub-sample of the full sample of simulated light
curves. The redshift on the x axis ($z_{\mathrm{sim}}$) is the initial adopted
redshift of the supernovae. The dotted lines indicate the 
$|z_{\mathrm{fit}}-z_{\mathrm{sim}}|/(1+z_{\mathrm{sim}})>0.15$ limits.}
\label{fig:zacc}
\end{figure}

The reference epoch $R$ and $I$ images were used together with the $UBV$ deep
images to obtain photometric redshifts for the supernova host galaxies. The
redshifts are calculated using the GOODSZ code \citep{2010ApJ...724..425D}.
This is a $\chi^2$ template fitting code that uses empirical spectral energy
distributions from \citet{1980ApJS...43..393C} and \citet{1996ApJ...467...38K}.
The photometric redshifts were found to have an accuracy of $\sigma_z=0.07$
where the redshift scatter $\Delta z$ is given by $\Delta z = \sigma_z (1+z)$.
Details on the photometric redshift technique will be presented in \citet{laia}. 
We require the galaxies to be detected in at least 3
filters to trust the resulting redshifts, something which is fulfilled for all
of the 16 supernovae.  The hosts were selected by choosing the closest galaxy
in physical distance, calculating the distance from the photometric redshift
and the angular distance. For two of the supernovae (SN309 and SN357) this
means that the closest galaxy in terms of angular distance is not the host
galaxy.  These galaxies are at much higher redshifts ($z\gtrsim 3$), thus it is
extremely unlikely we would be able to detect a supernova in them. More details
on the host galaxy identification is presented in \citet{2011A&A...532A..29M}.

The photometric redshifts, together with the 68\% confidence limits (as
estimated by the $\chi^2$ fitting), were used to construct a Bayesian prior for
the supernova typing.  The typing code then outputs a most likely redshift for
the best fitting supernova type for each supernova \citep[details are given
in][]{2011A&A...532A..29M}. This is the redshift estimate used for the
supernovae in the rate calculations.  Fig.~\ref{fig:zacc} shows the redshift
accuracy for the Monte Carlo simulated supernova sample described in
Sect.~\ref{sec:sample}. When typing the simulated supernovae we assume that the
redshift errors for the host galaxies are equal to the redshift scatter,
$\Delta z$. The resulting redshift accuracy for the SNe with host galaxy
redshifts is given by $\sigma_z=0.07$ with an outlier fraction (defined as
having $|z_{\mathrm{sim}}-z_{\mathrm{fit}}|>0.15$) of 3\% over the full range of
redshifts. The resulting redshift accuracy from the ``photometric redshift$+$SN
typing´´ scheme is thus unchanged from the pure photometric redshift scheme.   

For reference, we also show the accuracy for SNe typed
with a flat redshift prior (i.e. sources without any photometric redshift
information).  This is worse, with $\sigma_z =0.11$ and an outlier fraction of
4\%. For both populations it should be noted that the majority of the
outliers comes from the $z_{\mathrm{sim}}=0.2-0.4$ region and is due to mistyping of the
supernovae.

The typing was also rerun using a flat redshift prior for all of the discovered
SNe. The fitted redshifts from this run were all within the 68\% confidence
limit of the host photometric redshifts and none of the SNe changed main type.
Based on this the chosen hosts are thus likely to be the correct ones, although
it is certainly possible that some hosts have been misidentified. In any case,
neither the typing nor redshifts are appreciably affected by the possible host
misidentification. 

\section{Supernova rate determination}
\label{sec:method}
\begin{table} 
\caption{Properties of the supernova photometric templates}	     \label{table:templ} \centering \begin{tabular}{l c c c c c} \hline\hline Type & $M_B$ & $\sigma_M$ & Stretch &
Fraction\\ \hline Ia -- 91T    & -19.64 & 0.30 & 1.04 & 0.10 \\ 
Ia -- normal & -19.34 & 0.50 & 1.00 & 0.52\\ 
Ia -- faint  & -18.96 & 0.50 & 0.80 & 0.18 \\ 
Ia -- 91bg   & -17.84 & 0.50 & 0.49 & 0.20\\ 
\hline					 Ibc -- bright & -19.34 & 0.46 & N/A & 0.016\\ 
Ibc -- normal & -17.03 & 0.49 & N/A & 0.324 \\ 
IIL           & -17.23 & 0.38 & N/A & 0.061\\ 
IIn           & -18.82 & 0.92 & N/A & 0.051\\ 
IIP           & -16.66 & 1.12 & N/A & 0.547\\ 
\hline					 \end{tabular} 
\\
\tablefoot{$\langle M_B\rangle$ is the absolute magnitude in the
Johnson-B filter at peak, $\sigma_M$
is the dispersion in the peak magnitude. Fraction refers to the fraction within 
each main type in a volume limited sample.}\\
\end{table}
The supernova rates were determined using a Monte Carlo method
\citep{2004ApJ...613..189D}. Using a set of supernova templates (see
Table~\ref{table:templ}) we simulate a number of supernova light curves of
different types and at different redshifts with time of explosion set within
our detection window. The requirement that each SN has to be detected
in two consecutive epochs means that this window stretches from approximately 5
months before until five months after the start of monitoring. The main input
parameters are the intrinsic supernova rates (Ia and CC) for a given redshift
bin and volume-limited fractions of the different types. The template light
curves are calculated from absolute magnitude light curves \citep[mainly
from][]{2002AJ....123..745R, 2004ApJ...613..189D, 2006AJ....131.2233R} and a
set of spectra from \citet{2007NugentMISC}. For more information on how the
template light curves are built see \citet{2011A&A...532A..29M} and references
therein.

The volumetric fractions for the Ia subtypes are adapted from
\citet{2011MNRAS.tmp..413L}, treating their Ia-HV sub-category as Ia--normal
SNe. The Ia--faint category is used in the typing of supernovae to be able to
better characterise normal Ia supernovae with a somewhat lower value of
stretch. The fraction of low-stretch Ia SNe is estimated from the distribution
of stretch value in the Supernova Legacy Survey \citep{2006ApJ...648..868S} Ia
sample.  The core collapse fractions are based on a compilation of supernovae
from \citet{2009MNRAS.395.1409S} and \citet{2011MNRAS.tmp..413L}, treating IIb
SNe as Ibc--normal supernovae, since, with our time sampling of the light
curves, these subtypes will look very similar.

Each supernova is also given a host galaxy extinction. For type Ia SNe we use
the parametrisation of the \citet{2005MNRAS.362..671R} simulations presented in
\citet{2006AJ....132.1126N}, while for core collapse SNe we use Monte Carlo
simulations based on the models of \citet{2005MNRAS.362..671R}. The effects of
spiral arms and the bulge component are considered negligible in the core
collapse SN simulations and are not included. The extinction is scaled with the
$V$-band optical depth through a simulated face-on galaxy at zero radius with
$\tau_{V}(0)=2.5$, which provides a reasonable match to the observed host
galaxy extinction distribution for CC SNe within 12 Mpc
\citep{2012arXiv1206.1314M}. To calculate the wavelength dependence of the
extinction, we use a Cardelli extinction law \citep{1989ApJ...345..245C} with
$R_V=3.1$ for Type Ia SNe and a Calzetti attenuation law
\citep{2000ApJ...533..682C} with $R_V=4.05$ for core collapse SNe. In
Sect.~\ref{sec:systerrs}, we investigate how the choice of extinction models
and extinction laws affects the results.  It is important to note that this
extinction is the result of simulations of normal spiral/elliptical galaxies.
In this case the modelled mean extinction only goes above $A_V
\sim 1$ mag. for galaxies with inclination higher than 60 degrees. According
to the simulations, no supernovae in these normal galaxies exhibit very high
extinctions. But observations of local supernovae show that this is untrue,
e.g. SN~2002hh \citep{2006MNRAS.368.1169P} and SN~2009hd
\citep{2011ApJ...742....6E} both have host galaxy extinctions of $A_V \sim 4$
mag. The extinction adopted for our simulated supernova light curves thus only
include the effect of small to moderate levels of extinction. In
Sect.~\ref{sec:obsc} we describe how the output rates are de-biased to account
for the missing population of highly extinguished supernovae in normal galaxies
as well as in U/LIRGs.  

The light curves are then fed through the same detection procedure as
the real supernova light curves (see Sect.~\ref{sec:sample}).  It
should be noted that even with a quite conservative photometric limit
($3 \sigma$ in four data points) there are issues with completeness.  The
detection efficiency of the survey does affect the observed rates since
it starts to drop below one before the $3 \sigma$-limit is reached. The
detection efficiencies described below are thus used to give a
probability that a supernova is detected in a certain epoch and filter,
and are included in the Monte Carlo simulations. We have used the
detection efficiencies derived for hosts of intermediate brightness for
all supernovae. The effect of this choice on the rates is small (see
Sect.~\ref{sec:systerrs}).

The output number of detectable SNe from the simulations is then compared to
the observed number in the chosen redshift bin. Since the redshifts for the SNe
have quite high uncertainties (being based on photometric redshifts) we
distribute the SNe in the bins according to a Gaussian distribution with the
redshift uncertainty $\Delta z=0.07(1 + z)$. We denote the number of
redistributed SNe in each bin N$_{\mathrm{redist}}$. The simulations are then iterated
until the output number matches N$_{\mathrm{redist}}$.  At that point the input
intrinsic SN rate that produces the correct number of observed SNe is chosen as
the true supernova rate.
\subsection{Detection efficiencies} 
\begin{figure} 
\centering
\includegraphics[width=8cm]{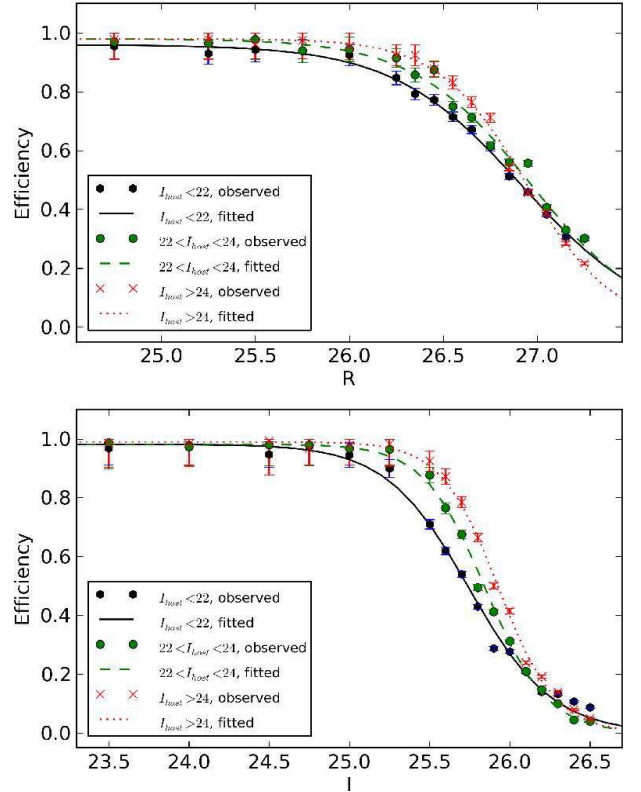} 
\caption{An example of the detection efficiency for a
given epoch for different host galaxy brightness. The efficiencies from
simulations of supernovae in the field are given by the points
while the lines are the S-curve fits to the efficiencies as described in
the text. Errors for the measured efficiencies are binomial.  Note that a
total of 30 (one for each of five epochs $\times$, two filters $\times$, 
and three host galaxy types) S-curve fits have been used to estimate the detection
efficiencies of the full survey. This figure is available in colour in
the electronic version of the article.}
\label{fig:deteff} 
\end{figure}
The detection efficiency for each epoch and filter was determined by
simulating supernovae at variable brightness in the actual search images
and then running the standard detection pipeline on the simulated
frames. The procedures used are described in
\citet{2008A&A...490..419M}. We placed supernovae in different
environments, i.e. with different host galaxy brightness, to study the
effect of background light on the efficiencies.  The detection
efficiencies are notably worse when the SNe are situated in bright
galaxies (defined as having $m_I<22$). In Fig.~\ref{fig:deteff}
a sample of the detection efficiency testing is presented, showing the
efficiencies for the three different host galaxy brightness modes. It
should be noted that the 50\% efficiency limits roughly corresponds to
the $3 \sigma$ rejection limits.

To use the efficiency curves in the Monte Carlo simulations, and to
smooth out possible outliers in the measured efficiency, we parametrise
the efficiencies using a S-curve parametrisation \citep[as previously used by,
e.g.][]{2004ApJ...613..200S} given by:
\begin{equation}
\label{eq:scurve}
\epsilon(m)=\frac{\epsilon_0}{1+e^{(m-m_\mathrm{c})/S}},
\end{equation}
where $\epsilon(m)$ is the fitted detection efficiency and $m$ the magnitude.
$\epsilon_0$ is the maximum efficiency (which is very close to one for most of
our fits), $m_\mathrm{c}$ the magnitude when the efficiency has dropped to 50 \%, and
$S$ a parameter that determines how fast the drop occurs. The best-fitting
parameters for each epoch, filter and host galaxy brightness are found with a
simple $\chi^2$ optimisation algorithm and are shown as lines in
Fig.~\ref{fig:deteff}

\subsection{De-biasing the rates for extreme host galaxy extinction}
\label{sec:obsc} 
The star formation in the local universe takes place mostly in galaxies with
low amounts of dust and thus the supernovae discovered in the local universe
often have low extinctions. The use of core collapse supernovae as tracers of
recent star formation is thus feasible at low redshifts
\citep[e.g.][]{2012A&A...537A.132B}. However, as we go to higher redshifts the
bulk of the star formation takes place in dusty galaxies with high infrared
luminosities (LIRGs and ULIRGs). A number of studies
\citep[e.g.][]{2005ApJ...632..169L,2009A&A...496...57M, 2011A&A...528A..35M}
have found the fraction of star formation taking place in LIRGs and ULIRGs to
increase rapidly towards $z\sim 1$, such that approximately half of the star
formation at $z\sim 1$ is taking place in these galaxies.  Furthermore, highly
extinguished supernovae are present also in normal spiral galaxies
\citep[e.g.][]{2012arXiv1206.1314M}. From now on we denote the effect these two
factors have on the supernova rates obscuration, to distinguish it from the
standard host galaxy extinction of light from SNe in normal spiral and
elliptical galaxies.

\citet{2007MNRAS.377.1229M} compiled the star formation densities for different
redshifts derived from UV and infrared observations in a number of studies.
They used these results together with their own estimates on how many SNe are
lost due to obscuration by dust in local starburst galaxies, LIRGs and ULIRGs
to derive a correction for supernova rates at high redshifts.  The estimates
were based on SN searches conducted in such galaxies by that time
\citep{2002A&A...389...84M,2003A&A...401..519M}, which had found few SNe.
Another caveat with the previous studies of this effect is that it was, at that
time, almost completely unknown what kind of galaxies the high redshift LIRGs
and ULIRGs were. They were selected based on high luminosities in the mid- and
far-infrared wavelengths, and the assumption made in
\citet{2007MNRAS.377.1229M} was that they were the same kind of starbursting,
irregular and compactly star forming galaxies as in the local universe. At the
time this was not an unreasonable assumption, but later developments (Spitzer
and Herschel observations in particular) have shown that the high redshift
U/LIRG population is dominated by disk galaxies with more constant, although high,
star formation rates \citep{2010ApJ...713..686D, 2011A&A...533A.119E,
2011arXiv1110.4057K}, the so called main sequence (MS) galaxies. These galaxies
have a higher content of gas and dust than local disk galaxies, but they do not
exhibit the kind of compact star formation found in local U/LIRGs. A smaller fraction
of the high redshift U/LIRGs are more similar to the local counterparts
\citep{2011A&A...533A.119E, 2011arXiv1110.4057K}.

\subsubsection{De-bias factors for the core collapse supernovae} 
For SNe in normal galaxies we have already taken into account the effect of low
to moderate amounts of extinction, but, as previously mentioned, there are
supernovae with extreme extinction also in normal disk galaxies.
\citet{2012arXiv1206.1314M} find that $15^{+21}_{-10}$\% of the supernovae in
these galaxies have extinctions significantly higher than predicted by
simulations following the recipe in \citet{2005MNRAS.362..671R}.  These are SNe
that would likely be missed in dusty regions of the normal galaxies by
magnitude limited surveys. In the local universe almost all U/LIRGs are also
characterized as starburst galaxies \citep[e.g.][]{2010ApJS..188..447P}. In
these galaxies less than $\sim 20$\% of the CC SNe can be detected in optical
searches and in some extreme cases, such as Arp~220, the entire SN population
is obscured by the large amounts of dust and can only be detected in radio
\citep[e.g.][]{2007ApJ...659..314P}. \citet{2012arXiv1206.1314M}
investigated the SN population of the nearby LIRG Arp 299 and found that
$83^{+9}_{-15}$\% of SNe exploding in this galaxy will remain undetected by
optical SN searches. They then assume that Arp~299 is representative of compact
starbursting U/LIRGs and adopt this figure as the missing fraction of
supernovae in these galaxies. The missing fraction of SNe in high redshift
U/LIRGs is highly uncertain. For the main sequence (i.e.  non-starburst)
galaxies little is known about the extinction, although they have been found to
be disk-like and not as compact as their local counterparts
\citep{2011arXiv1110.4057K}. \citet{2012arXiv1206.1314M} assume a missing
fraction of $37^{+46}_{-18}$ \% for these galaxies (see
Sect.~\ref{sec:systerrs} for a discussion on the systematic errors resulting
from this decision).

\citet{2012arXiv1206.1314M} calculate the de-bias factor as a function of
redshift by using the relative contributions to the cosmic star formation
density of normal galaxies (defined as galaxies with $L_{\mathrm{IR}}<10^{11} L_\odot$),
LIRGs, and ULIRGs from \citet{2011A&A...528A..35M} and the relative
contributions of the compact and main sequence channels from
\citet{2011arXiv1110.4057K}. Furthermore, they assume that Arp~299 only
represents the U/LIRGs that are starbursting, i.e. lie more than three times above
the specific star formation rate (sSFR) MS locus at high redshift. 

We adopt the de-biasing factors for the CC SN rates in the two redshift bins
under consideration from \citet{2012arXiv1206.1314M}. The final de-bias
factors to correct the CC SN rates for obscuration in our two redshift bins is
1.34 at $0.1<z<0.5$ and 1.52 at $0.5\leq z<0.9$. In
Sect.~\ref{sec:systerrs} we give the errors on these factors and discuss the
systematic errors on the rates resulting from the calculation.

\subsubsection{De-bias factors for the type Ia supernovae}  
\label{sec:debiasIa}
The previous section dealt with de-biasing the CC SN rate, but it should be
noted that also Ia supernovae may be missed in SN surveys due to high
extinction in star forming galaxies. Multiple studies have shown that the rate
of Ia SNe is correlated with the current star formation rate in the host
galaxies \citep{2006MNRAS.370..773M,2006ApJ...648..868S}. This indicates that
there is a so called prompt channel of thermonuclear explosion
\citep{2005ApJ...629L..85S}, or that the delay time distribution extends to
time intervals as short as $\sim$420 Myrs \citep{2011MNRAS.tmp..307M}.

No Ia SN has ever been detected in the compact star forming cores of starburst
galaxies. But, given that the extinction is extreme in such environments and
that Ia SNe are faint in radio, this may be the result of selection
effects. However, the delay time distribution of Ia SNe has so far only been
determined down to a time difference of 420 Myrs \citep{2011MNRAS.tmp..307M},
while the starburst phase in U/LIRG starbursts normally has a duration of $\sim
100$ Myrs \citep[e.g.]{2006A&A...458..369M}. After this time the starburst normally
enters the post-starburst phase where the lack of gas makes the star formation
stop \citep[e.g.][]{2007MNRAS.377.1222G}. Most of the dense dust and
gas shroud in a starburst core will likely have been disrupted and blown off in
the post-starburst phase \citep[e.g.][]{2007ApJ...663L..77T}, and any Ia SNe
exploding after a delay time of $\gtrsim 100$ Myrs will thus not experience
higher extinction than a supernova in a normal star forming galaxy. 
  
Similar to what has been found for CC SNe, also a small fraction of the Ia SNe
will be subject of extreme extinction in normal star forming disk galaxies.
This fraction will depend on the delay time distribution and is likely
significantly lower than the fractions found for CC SNe. At higher
redshifts a fraction of type Ia SNe will explode in the main
sequence channel of the U/LIRG population. These SNe will, as the CC SNe,
suffer from higher extinction than the normal extinction models predict,
but no estimates on this effect can be found in the literature and a detailed
study is outside the scope of this paper.

The fraction of Ia SNe missed at high redshift has been estimated by
\citet{2007MNRAS.377.1229M}, but suffers from the same problems as the core
collapse estimate in addition to systematic errors from the assumptions on
delay time distribution and starburst lifetime (few details on the
calculations are given in that paper) as described above. Because of the added
uncertainties inherent in estimating a missing fraction for type Ia SNe we do
not try to compute any missing fraction for the Ia SNe. Furthermore, it should
be noted that any assumptions on the delay time distribution may introduce a
circular argument. To limit the circularity, and to avoid using a de-bias
factor for which the systematic error is unknown, we assume that no 
Ia SNe will be missed because of obscuration (but note that we do apply a
correction for the normal host galaxy extinction). In Section~\ref{sec:systerrs} we
estimate the systematic error resulting from this assumption.

\section{Results}
\label{sec:results}
\begin{table*}
\caption{SVISS supernova rates}	     \label{table:rates} \centering \renewcommand{\arraystretch}{1.3} \begin{tabular}{c c c} \hline\hline \multicolumn{3}{c}{Ia Supernovae}\\
\hline
$z$              & & $0.3<z<0.8$\\
$\langle z\rangle$& & 0.62 \\
N$_{\mathrm{raw}}$ & & 5\\
N$_{\mathrm{redist}}$ & & 4.92\\
R$_{\mathrm{Ia}} \times 10^{-4}$(no ext.)& &1.08$_{-0.49\,-0.24}^{+0.69\,+0.23}$\\
\hline
R$_{\mathrm{Ia}} \times 10^{-4}$(ext.)& & 1.29$_{-0.57\,-0.28}^{+0.88\,+0.27}$\\
\hline
\multicolumn{3}{c}{CC Supernovae}\\ 
\hline		
$z$ & $0.1<z<0.5$ & $0.5\leq z<0.9$\\
$\langle z\rangle$& 0.39 & 0.73 \\
N$_{\mathrm{raw}}$ & 3 & 5\\
N$_{\mathrm{redist}}$ & 3.28 & 3.27\\
R$_{\mathrm{CC}} \times 10^{-4}$(no ext.)&$1.83_{-0.98\,-0.79}^{+1.72\,+0.97}$ &
$2.59_{-1.29\,-0.80}^{+2.14\,+1.17}$\\
R$_{\mathrm{CC}} \times 10^{-4}$(ext.)&$2.46_{-1.33\,-1.42}^{+2.30\,+1.75}$ &
$4.21_{-2.05\,-1.98}^{+3.49\,+2.94}$ \\
\hline
R$_{\mathrm{CC}} \times 10^{-4}$(ext.+obsc.)& $3.29_{-1.78\,-1.45}^{+3.08\,+1.98}$ &
$6.40_{-3.12\,-2.11}^{+5.30\,+3.65}$ \\
\hline
\end{tabular} 
\renewcommand{\arraystretch}{1} \tablefoot{$\langle z\rangle$ is the volume averaged redshift over the
given redshift range.  The supernova rates R$_{\mathrm{Ia/CC}}$ are in units of
yr$^{-1}$ Mpc$^{-3}$ h$^3_{70}$ and given with and without corrections
for extinction and obscuration. N$_{\mathrm{raw}}$ is the raw number of
supernovae per bin and N$_{\mathrm{redist}}$ is the number when taking the
redshift uncertainty into account.}
\end{table*}
\begin{table*}
\caption{Supernova rates in the literature}	     \label{tab:littrates} \centering \renewcommand{\arraystretch}{1.3} \begin{tabular}{l l l} \hline\hline $\langle z\rangle$ & R$_{\mathrm{Ia/CC}} \times 10^{-4}$ 
& Reference\\ 
\hline 
\multicolumn{3}{c}{Ia Supernovae}\\
\hline
$<0.0066$ ($<28$ Mpc)& $\geq 0.35$ & Smartt et al. (2009)$^*$\\
0.01 & $0.24^{+0.07}_{-0.07}$ & Cappellaro et al. (1999)\\
$<0.014$ ($<60$ Mpc)& $0.265^{+0.034}_{-0.034}$($^{+0.043}_{-0.043}$) & Li et al. (2011)\\
0.0375   & $0.278^{+0.112}_{-0.08}$($^{+0.015}_{-0.0}$) & Dilday et al. (2010)\\
0.1      & $0.259^{+0.052}_{-0.044}$($^{+0.018}_{-0.001}$) & Dilday et al. (2010)\\
0.15     & $0.307^{+0.038}_{-0.034}$($^{+0.035}_{-0.005}$) & Dilday et al. (2010)\\
0.15     & $0.32^{+0.23}_{-0.23}$($^{+0.07}_{-0.06}$) & Rodney \& Tonry (2010)\\
0.2      & $0.348^{+0.032}_{-0.030}$($^{+0.035}_{-0.007}$) & Dilday et al. (2010)\\
0.25     & $0.365^{+0.031}_{-0.028}$($^{+0.182}_{-0.012}$) & Dilday et al. (2010)\\
0.3      & $0.434^{+0.037}_{-0.034}$($^{+0.396}_{-0.016}$) & Dilday et al. (2010)\\
0.3      & $0.34^{+0.16}_{-0.15}$($^{+0.21}_{-0.22}$) & Botticella (2008)\\
0.35     & $0.34^{+0.19}_{-0.19}$($^{+0.07}_{-0.03}$) & Rodney \& Tonry (2010)\\
0.442    & $0.00^{+0.50}_{-0.00}$($^{+0.00}_{-0.00}$) & Barbary et al. (2012)\\
0.45     & $0.31^{+0.15}_{-0.15}$($^{+0.12}_{-0.04}$) & Rodney \& Tonry (2010)\\
0.47	 & $0.42^{+0.06}_{-0.06}$($^{+0.13}_{-0.09}$) & Neill et al. (2006)\\
0.47     & $0.80^{+0.37}_{-0.27}$($^{+1.66}_{-0.26}$) & Dahlen et al. (2008)\\
0.55     & $0.54^{+0.099}_{-0.086}$($^{+0.11}_{-0.116}$) & Pain et al. (2002)$^*$\\
0.55     & $0.32^{+0.14}_{-0.14}$($^{+0.07}_{-0.07}$) & Rodney \& Tonry (2010)\\
\textbf{0.62}& \textbf{1.29$^{+0.88}_{-0.57}$($^{+0.27}_{-0.28}$)} & \textbf{SVISS (this work)}\\
0.65     & $0.49^{+0.17}_{-0.17}$($^{+0.14}_{-0.08}$) & Rodney \& Tonry (2010)\\
0.74     & $0.79^{+0.33}_{-0.41}$ & Graur et al. (2011), errors include systematics\\
0.75     & $0.68^{+0.21}_{-0.21}$($^{+0.23}_{-0.14}$) & Rodney \& Tonry (2010)\\
0.807    & $1.18^{+0.60}_{-0.45}$($^{+0.44}_{-0.28}$) & Barbary et al. (2012)\\
0.83     & $1.30^{+0.33}_{-0.27}$($^{+0.73}_{-0.51}$) & Dahlen et al. (2008)\\
0.85     & $0.78^{+0.22}_{-0.22}$($^{+0.31}_{-0.16}$) & Rodney \& Tonry (2010)\\
0.95     & $0.76^{+0.25}_{-0.25}$($^{+0.32}_{-0.26}$) & Rodney \& Tonry (2010)\\
1.05     & $0.79^{+0.28}_{-0.28}$($^{+0.36}_{-0.41}$) & Rodney \& Tonry (2010)\\
1.187    & $1.33^{+0.65}_{-0.49}$($^{+0.69}_{-0.26}$) & Barbary et al. (2012)\\
1.21     & $1.32^{+0.36}_{-0.29}$($^{+0.38}_{-0.32}$) & Dahlen et al. (2008)\\
1.23     & $0.84^{+0.25}_{-0.28}$ & Graur et al. (2011), errors include systematics\\
1.535    & $0.77^{+1.07}_{-0.54}$($^{+0.44}_{-0.77}$) & Barbary et al. (2012)\\
1.61     & $0.42^{+0.39}_{-0.23}$($^{+0.19}_{-0.14}$) & Dahlen et al. (2008)\\
1.69     & $1.02^{+0.54}_{-0.37}$ & Graur et al. (2011), errors include systematics\\
\hline
\multicolumn{3}{c}{CC Supernovae}\\ 
\hline		
$\lesssim 0.003$ ($<11$ Mpc)& $\geq 1.6^{+0.4}_{-0.4}$ & Botticella et al. (2012)$^*$, stat. errors only \\
$\sim 0.003$ (6-15 Mpc)& $\geq 1.5^{+0.4}_{-0.3}$ & Mattila et al. (2012)$^*$, stat. errors only \\
$<0.0066$ ($<28$ Mpc)& $\geq 0.88$ & Smartt et al. (2009)$^*$\\
0.01 & $0.58^{+0.19}_{-0.19}$ & Cappellaro et al. (1999)\\
$<0.014$ ($<60$ Mpc)& $0.62^{+0.067}_{-0.067}$($^{+0.17}_{-0.15}$) & Li et al. (2011) \\
0.21 & $1.148^{+0.43}_{-0.34}$($^{+0.42}_{-0.36}$ & Botticella et al. (2008)\\
0.3  & $1.42^{+0.3}_{-0.3}$($^{+0.32}_{-0.24}$) & Bazin et al. (2009)\\
0.3  & $2.51^{+0.88}_{-0.75}$($^{+0.75}_{-1.86}$) & Dahlen et al. (2004)\\
\textbf{0.39}& \textbf{$3.29^{+3.08}_{-1.78}$($^{+1.98}_{-1.45}$)} & \textbf{SVISS (this work)}\\
0.66 & $6.9^{+9.9}_{-5.4}$ & Graur et al. (2011), errors include systematics \\ 
0.7  & $3.96^{+1.03}_{-1.06}$($^{+1.92}_{-2.6}$) & Dahlen et al. (2004)\\
\textbf{0.73}& \textbf{$6.40^{+5.30}_{-3.12}$($^{+3.65}_{-2.11}$)}& \textbf{SVISS (this work)}\\
\hline
\end{tabular} 
\renewcommand{\arraystretch}{1} \tablefoot{$\langle z\rangle$ is the average redshift for the redshift
interval, as given in the respective paper. The supernova rates R$_{\mathrm{Ia/CC}}$ are
in units of yr$^{-1}$ Mpc$^{-3}$ h$^3_{70}$. Errors are statistical
(systematical) unless otherwise noted. The rates are corrected for extinction
given in the original reference. In a few cases (references marked with $^*$)
no extinction correction have been done. The local CC rates (papers by
Botticella and Mattila) have been measured in small volumes and are likely
dominated by a local overdensity of star formation.}
\end{table*}
\subsection{Core collapse supernova rates}
\begin{figure*} 
\centering
\includegraphics[width=16cm]{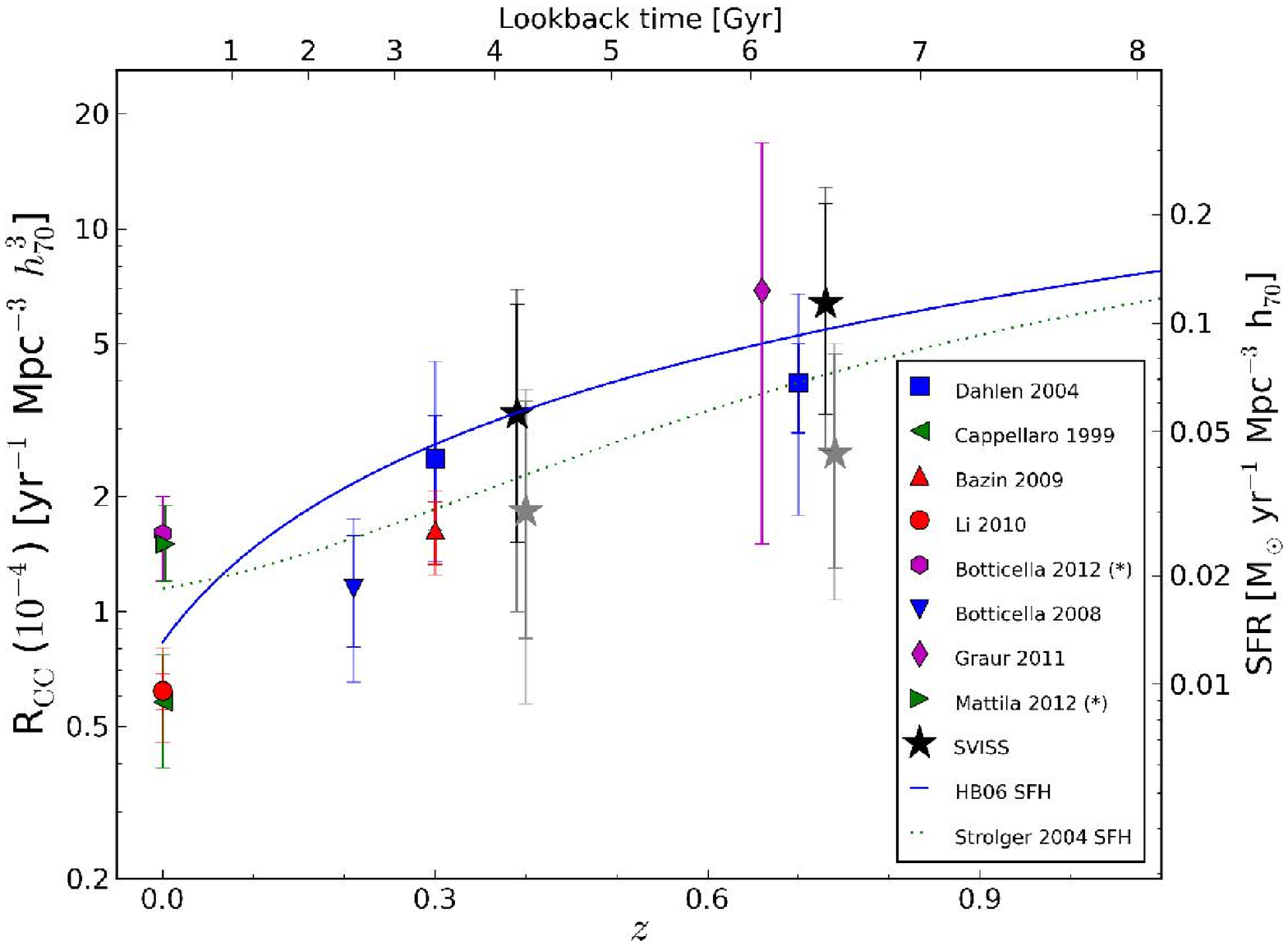}
\caption{Core collapse supernova rates determined from the SVISS SNe (black
stars). Grey stars are displaced by $+0.01$ in $z$ and show the SVISS rates
without extinction correction. Also shown is a collection of measured rates
from other authors. We plot star formation histories from two different
sources, scaled to supernova rates through the use of Eq.~\ref{eq:snr}. Error
bars are statistical with total errors (statistical and systematic added in
quadrature) as a transparent/faded extra error bar for all surveys. Redshift
bin sizes are not shown, but are given in Table~\ref{table:rates}. The rate
from \citet{1999A&A...351..459C} has been rescaled from SNu to a volumetric
rate by assuming a local B-band luminosity density $2.0\times 10^8$ h$_70$
$L_{\sun}$ Mpc$^{-3}$ at redshift $\sim 0$. The local CCSN rates from
\citet{2012A&A...537A.132B} and \citet{2012arXiv1206.1314M} have been
measured in small volumes and are likely dominated by inhomogeneity. The star
formation rate in the local volume is higher than the average over an extended
volume and can therefore cause the SNR to be higher than expected.}
\label{fig:ccrates} 
\end{figure*}

The core collapse supernova rate in two redshift bins (0.1--0.5, 0.5--0.9) is
shown in Fig.~\ref{fig:ccrates} and in Table~\ref{table:rates}. 
The two bins do not include all of the SNe in the sample. By binning
differently the total number can be increased, but this comes at the price of
higher statistical uncertainty (with more bins) and higher systematic
uncertainties (with wider bins).
Both the rates corrected for extinction/obscuration and the raw rates
are shown in the table but the figure show the corrected values. The two sets
of error bars in Fig.~\ref{fig:ccrates} show the statistical errors, and
statistical and systematic errors added in quadrature, unless otherwise noted.

In Table~\ref{tab:littrates} and Fig.~\ref{fig:ccrates} we show a comparison of
our rates with rates reported by other authors.  At low redshift we plot the
rate estimates of LOSS \citep{2011MNRAS.tmp..317L}, which is the largest SN
survey to date in the local volume with a total number of core collapse SNe of
440. We also plot the low redshift rates determined by
\citet{1999A&A...351..459C}, \citet{2012A&A...537A.132B}, and
\citet{2012arXiv1206.1314M}.  At slightly higher redshifts we plot the results
from \citet{2008A&A...479...49B}, which also includes the data presented in
\citet{2005A&A...430...83C}. The second survey with rates based on a large
number of CC SNe (117 SNe) is the Supernova Legacy Survey
\citep[SNLS,][]{2009A&A...499..653B} which provides a data point with small
error at $z=0.3$. None of these surveys include de-biasing for obscuration.

At higher redshifts we plot the rates from the GOODs determined by
\citet{2004ApJ...613..189D}, which includes the effect of extinction in normal
spiral galaxies, but not de-biasing for obscuration and the recent measurements
from the Subaru Deep Field \citep{2011MNRAS.tmp.1508G}, which includes
de-biasing for obscuration and extinction in normal galaxies. For comparison,
the figure also shows estimates of the star formation history from two
different sources, \citet{2004ApJ...613..200S}  and
\citet{2006ApJ...651..142H}, scaled to supernova rates (see
Sect.~\ref{sec:sfr}). Both are corrected for extinction.

The rate at $z=0.39$ is consistent with the rates reported by other authors,
even though different approaches to extinction correction is used.  Our CC rate
at high redshift agrees well with the observations of
\citet{2004ApJ...613..189D} and \citet{2011MNRAS.tmp.1508G}.
For a discussion on how our supernova rate measurements compares to the star
formation history, see Sect.~\ref{sec:disc}.

\subsection{Ia supernova rates}
\begin{figure*} 
\centering
\includegraphics[width=16cm]{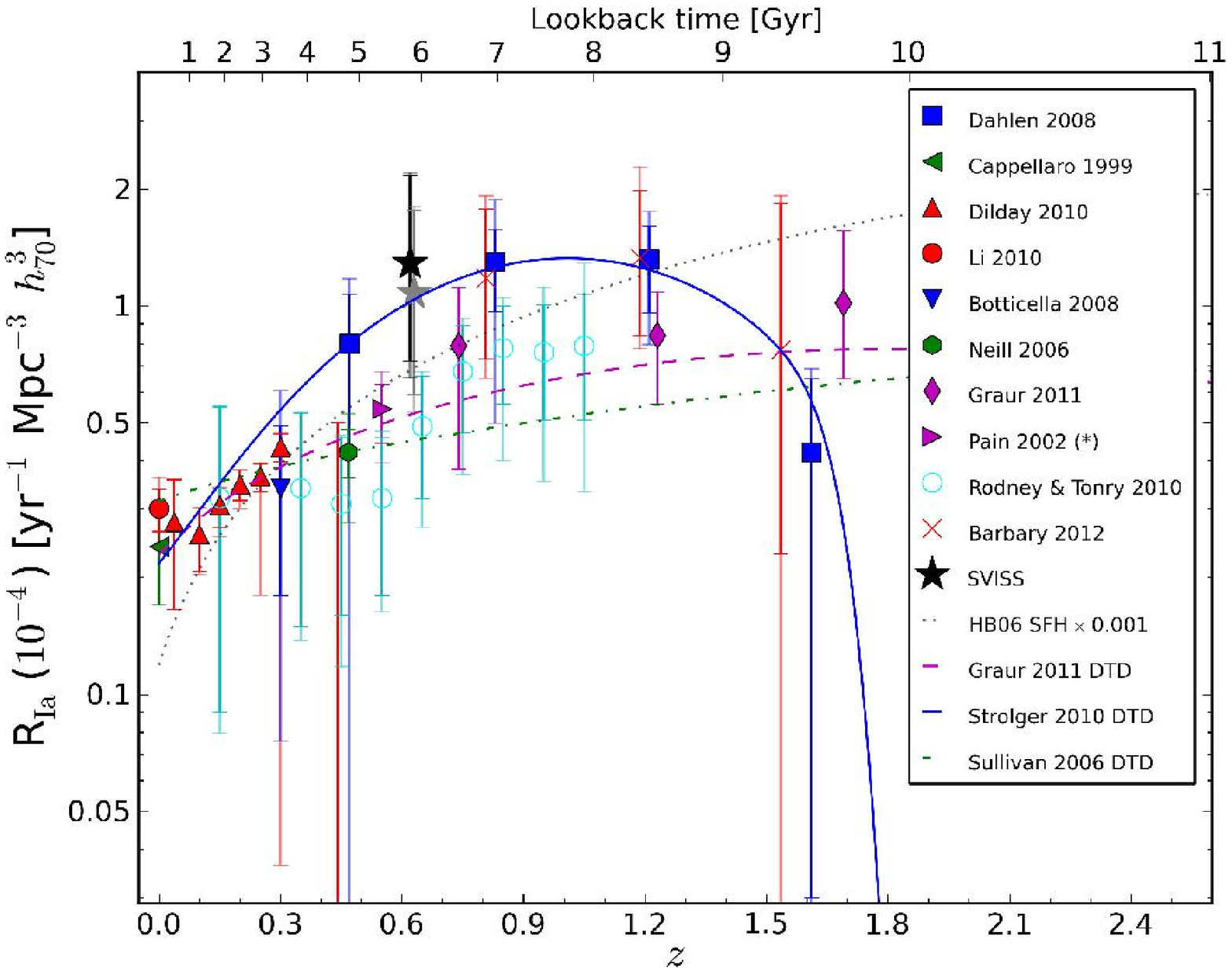} 
\caption{Type Ia supernova rates determined from the SVISS SNe (black star).
The grey star is displaced by $+0.01$ in $z$ and show the rate without
extinction correction.  Also shown is a collection of published rates from the
literature. We plot SN Ia rates resulting from assumptions on the delay time
distribution combined with the star formation history. Error bars are
statistical with total errors (statistical and systematic added in quadrature)
as a transparent/faded extra error bar for all surveys. Redshift bin sizes are
not shown, but are given in Table~\ref{table:rates}. The rate from
\citet{1999A&A...351..459C} has been rescaled from SNu to a volumetric rate by
assuming a local B-band luminosity density $2.0\times 10^8$ h$_{70}$ $L_{/sun}$
Mpc$^{-3}$ at redshift $\sim 0$.}
\label{fig:Iarates} 
\end{figure*}

The Ia supernova rate in one redshift bin (0.3--0.8) is shown in
Fig.~\ref{fig:Iarates} and in Table~\ref{table:rates}. The rates in the table
are given with and without extinction correction (no de-biasing to account for
obscuration have been done on the Ia rate). The choice of using one bin for the
Ia SNe is motivated by the low number of supernovae found at both low ($z<0.3$)
and high ($z>0.8$) redshifts. By using the redshift bin given above we minimise
the statistical errors. It should also be noted that the SNe outside the bin
have not been excluded from the rate calculation.  They contribute to the
N$_{\mathrm{redist}}$ of the bin, because they have redshift probability distributions
that stretches into the redshift interval. 

In Table~\ref{tab:littrates} and Fig.~\ref{fig:Iarates} we show a comparison of
our rate with rates reported by other authors.  At low $z$ we show the Ia rates
from the LOSS \citep{2011MNRAS.tmp..317L} and \citet{1999A&A...351..459C}. At
slightly higher $z$ we compare with the rates calculated from the SDSS
supernova search \citep{2010ApJ...713.1026D} and with the results from the
STRESS \citep{2008A&A...479...49B}. \citet{2010ApJ...723...47R} presented
revised Ia rates from the IfA Deep Survey out to $z=1.05$ using new techniques,
lower than those previously reported. The SNLS measured the Ia rate at $z=0.47$
using a large sample of supernovae, at similar redshift
\citet{2002ApJ...577..120P} presented some of the first rates determined from
cosmological survey data.  At higher redshifts \citet{2008ApJ...681..462D}
determined Ia rates from the GOODS (extending their sample from 2004).
\citet{2012ApJ...745...31B} has measured the volumetric SN Ia rate in the
background galaxies of the Hubble Space Telescope Cluster Supernova Survey.
\citet{2011MNRAS.tmp.1508G} reported the findings of the supernova search in
the Subaru Deep Field, their rate measurements are the highest redshift
measurements available and reach $z\sim 1.7$. 

Our rate measurement is higher than the results from many other surveys at this
redshift, but marginally consistent with all other measurements when taking
statistical and systematic errors into account.  While the uncertainty of our
rate measurement is too high and the redshift coverage too small to allow for
detailed fitting of different DTD models, we can compare our rates to models
introduced by other authors
\citep{2006ApJ...648..868S,2006AJ....132.1126N,2010ApJ...713...32S,
2011MNRAS.tmp.1508G} together with their chosen star formation history (see
Sect.~\ref{sec:dtd} for details). 

\subsection{Analysis of errors for the supernova rates}
\label{sec:systerrs}
The statistical errors are calculated using the prescription of
\citet{1986ApJ...303..336G}. The redshift bins have been chosen to
provide a reasonable number of sources in each bin to get similar
statistical errors in each bin.

We now proceed to study the systematic errors of our rate estimates. Given that
the total number of SNe is quite low, the statistical errors are high. One of
the goals of this study is thus to find out whether any of the systematic
effects can introduce errors larger than these. Calculating the mean
statistical relative error (mean of the upper and lower limit differences) for
the two types and all redshift bins we arrive at the following: 74\% for CC at
low $z$; 67\% for CC at high $z$; and 56\% for the Ia bin.  These can be
compared to the relative systematic errors calculated below. The second goal of
the study is to make an extensive list of systematic effects that can affect
supernova rate measurements and to try and estimate them. These effects will be
very similar for all supernova rate measurements of similar type and our
compilation can thus be of use in future surveys which may be limited by
systematic rather than statistical errors. A thorough study of the systematic
effects is also of great help when trying to find better observational
strategies in future surveys.

A compilation of the systematic errors can be found in 
Table~\ref{table:systerrs}.
\paragraph{Misclassification errors}
The systematic errors due to misclassification in our sample are given in 
\citet{2011A&A...532A..29M} and are on the order of 5--10 \%.
However, these estimates are really only valid when the number of
detected SNe is large enough.  For the small number statistic estimates
presented here the systematic errors have to be estimated in a different
way. The expected numbers of misclassified SNe in each type and redshift
bin is below one in all cases. To get a conservative estimate of the
misclassification systematic error we thus vary the observed number of
SNe by one (positive and negative) and recalculate the rates. This test
gives an error of approximately 20\%, but is dependent on the actual
number of sources in the given bin (a low source count gives a higher
error). For bins with more than two SNe the error is smaller than
the statistical errors.

The systematic misclassification errors are overestimated when computed
with this method, on average one SNe in our sample will have been
misclassified, not one per bin and type as the estimate indicates. But
it is impossible to say which bin and type that is affected by
misclassification, hence we give this error estimate.
\paragraph{Redshift}
The redshifts of the supernovae have uncertainties on the order of 0.1
when using the host photometric redshifts as a prior for the typing.
This leads to a redistribution in redshift which affects the rates. To
study this effect we redistribute the simulated SNe according to a
Gaussian distribution assuming $\sigma=0.07(1+z)$ and recalculate the
rates. This test is repeated 10,000 times to find the spread in the
rates. The systematic error varies between $\sim$15 to $\sim$30\% for
the different types and redshift bins, lower than the statistical errors
in all cases.
\paragraph{Detection efficiencies}
We determine the systematic errors from the detection efficiency
assumptions using the faint and bright host galaxy detection
efficiency estimates. The binomial errors of the efficiencies are
smaller than the difference between the results for the different host
types in almost all cases. The effect on the rates is on the order of
$\pm 2$\%, significantly lower than the statistical errors.
\paragraph{Photometric errors}
The SVISS photometric zero point calibration is accurate to within 
$\sim$ 5\%. Since we apply an absolute $3\sigma$ detection limit 
on our sample (see Sect.~\ref{sec:sample}) the estimated rates 
will be affected by a slight shift in the photometric zero points.
This effect will in reality be random over the two filters 
and seven epochs, to get a conservative estimate we vary the detection limits
by 0.1 magnitude. The resulting rates vary by $\sim$ 5\%.
\paragraph{Template choices}
Our selection of subtype light curve templates and their assumed
fractions influence the rates as well. The choices and assumptions we
have made are based on observational results
\citep{2002AJ....123..745R,2006AJ....131.2233R,2011MNRAS.tmp..413L}.
The fractions are subject to statistical uncertainties as well as
systematic uncertainties (e.g., evolution with redshift, selective
obscuration). There is also the possibility that very faint supernovae
are under-represented in magnitude-limited SN surveys
\citep{2011ApJ...738..154H}.

For the CC SNe we study the effect of introducing an additional light
curve template, a IIP template with a peak $M_B$ of $-14.39$ (as for the
faint core collapse SN~1987A) with fraction that changes the percentage
of CC SNe with $M_B>-15$ from 7\% to 30\%. The effect of this change on
the rates is quite large, it makes the rates go up by approximately
30\%. This is smaller than the statistical errors but could
help to explain any possible mismatch with the star formation history (as
identified by \citealt{2011ApJ...738..154H}). Varying the fractions
randomly have a very small effect on the rates since the changes tend to
cancel out.

For the Ia SNe we have made tests with different combinations of using or
not using the peculiar (91T and 91bg-like) and non-standard (Ia--faint) 
templates. We get the largest positive change in the rates when putting
the fraction of 91T SNe to zero (treating them as normal SN Ia's), on the order
of 1--2\%. The largest negative change is found when setting the fraction of
faint Ia supernovae (91bg and the Ia--faint subtypes) to zero, treating them as 
Ia--normal SNe, which results in a change of 10--15\% in the rates. These errors
should be considered to be the extreme limits, since there is plenty of
evidence that these faint Ia SNe do exist.
\paragraph{Peak magnitudes}
As described above the choice of template distribution also affects the
distribution of peak magnitudes for the simulated SNe. However, the
average peak magnitudes used in this work have associated uncertainties
(note that this is not the same as the scatter).

For the CC SNe the template peak magnitudes have uncertainties given by
\citet{2002AJ....123..745R} and \citet{2006AJ....131.2233R}. As
conservative estimate we allow the peak magnitudes of CC SNe to vary by
the 1$\sigma$ errors given in these papers and recompute the rates. The
resulting rates change by approximately 10\%, which is smaller than the
statistical errors.  

For the Ia SNe the peak magnitudes are much better constrained (due to their
use in precision cosmology). Assuming an uncertainty of 0.05 magnitudes for
the Ia subtype and 0.1 for the 91bg/91T subtypes we calculate the impact on 
the Ia rates. The effect is small, on the order of 3\%, much smaller than
the statistical errors.
\paragraph{Host galaxy extinction}
The effect of host galaxy extinctions are based on models by
\citet{2005MNRAS.362..671R} for type Ia
and core collapse SNe. For the type Ia SNe these models were used as given, while a different scaling was used for the CC SNe (see Sect.~\ref{sec:method}).
Using instead a positive Gaussian
distribution of $E(B-V)$ with $\sigma E(B-V)$=0.2 mag. as in
\citet{2006AJ....132.1126N} for Type Ia only changes the rate by $\sim
2$\%. For core collapse SNe, we have examined the effect of
applying an extinction model following \citet{2005MNRAS.362..671R} as given.
This has marginal effect on the rate in the low redshift bin (1\%)
and may affect the high redshift bin by 10\%.

Using the estimates from \citet{1998ApJ...502..177H} change the resulting rates
by a very small amount ($<1$\%). We also study the effect of using different
extinction laws for the extinction. For the CC SNe we use a Cardelli/$R_V=3.1$
law instead of the Calzetti law and for the Ia we use a steeper Cardelli law
with $R_V=2.1$ \citep[as advocated by][]{2008ApJ...686L.103G}. The impact on
the rates is small, on the order of 1\%.
\paragraph{Dust obscuration in the host galaxy}
\citet{2012arXiv1206.1314M} estimate lower and upper limits on the de-bias
factors for CC SNe from different assumptions on the amount of obscuration in
the SN host galaxies. The systematic errors on the CC SN rate resulting from
these limits are: $-10$\%/$+28$\% at $z=0.39$, and
$-10$\%/$+33$\% at $z=0.73$.  The upper and lower limits on the
de-bias factors corresponds to the upper and lower limits on the rates.

For the Ia SNe we assumed a nominal de-bias factor of 1.0 (i.e. a missing
fraction of zero). As an upper limit to this value we have chosen to use a de-bias
factor calculated from the formula given in \citet{2007MNRAS.377.1229M} for the
redshift in question. The resulting systematic error on the Ia rates is then
+8\% at $z=0.62$.
\paragraph{AGN contamination}
We used a control epoch with observations obtained one year after our
search period had ended to check whether any of our supernova candidates
showed variability over this longer period. Except for the possibility
of very rare peculiar SNe none of the SN subtypes should be detected one
year later.  The list of variable sources originally contained 31
entries, about ten of which showed variability over a long time-scale and
were excluded from the final list of SNe.  This routine is enough to
limit the amount of AGN contamination to very low levels. We do not give
this error in Table~\ref{table:systerrs} since we estimate it to be
smaller than all other systematic effects.
\paragraph{Cosmic variance}
The size of the SVISS field is fairly small ($4\times$56 sq. arcmin.).
Cosmic variance is therefore a possible uncertainty in the rate estimates. We
have estimated the cosmic variance by using the work of
\citet{2008ApJ...676..767T} \footnote{web based calculator available at
\url{http://casa.colorado.edu/~trenti/CosmicVariance.html}}, and assuming that the variance of SNe follow that of the overall galaxy population. The relative
cosmic variance for the type Ia SNe redshift bin is then 9\%. For the CC SNe low
redshift bin we estimate a relative variance of 12\%, and 10\% for the high
redshift bin.

\vspace{0.3cm}
In Table~\ref{table:systerrs} we also give the sum of all the systematic
errors (added in quadrature). These are also the systematic error estimates given in
Table~\ref{table:rates}. The co-added systematic errors are all smaller than the
statistical errors. The major contributions to the
systematic errors comes from misclassification, template
choices redshift uncertainties, and the obscuration correction.
\begin{table*}
\caption{Systematic errors}	     \label{table:systerrs} \centering \begin{tabular}{l c c c c} \hline\hline Error &\multicolumn{2}{c}{CC Supernovae} & \multicolumn{2}{c}{Ia Supernovae}\\
\hline 
 & $0.1<z<0.5$ & $0.5<z<0.9$ & $0.3<z<0.8$\\
\hline
Misclassification &25\% & 18\% & 16\% \\
Redshift & 32\% & 16\% & 9\% \\
Detection eff. & 2\% & 3\% & 1\% \\
Photometric & 4\% & 9\% & 4\% \\
Template choices & +31\% & +32\% & -12/+2\% \\
Peak magnitudes & 7\% & 15\% & 2\% \\
Extinction & +1\% & +10\% & 2\%\\
Obscuration & -10/+28\% & -10/+33\% & +8\% \\
Cosmic variance &12\% & 10\% & 9\% \\
\hline
Total systematic & -44/+60\% & -33/+57\% & -23/+24\%\\
Mean statistical & 74\% & 67\% & 56\% \\
\hline
\end{tabular} 
\tablefoot{Errors are two-sided unless the sign is given. The total errors have been computed by co-adding the individual errors in quadrature.}
\end{table*}

\subsection{Comparing the CC supernova rate to the cosmic star
formation history}
\label{sec:sfr}
The core collapse supernova rate can be compared to the cosmic star
formation history (SFH), since the time-scale for CC progenitors to explode is
much shorter than the cosmic time-scale.  The supernova rate is derived
from the star formation history by assuming an initial mass function
(IMF, denoted by $\xi$ below) and the mass range of the SN progenitors
($M_{\mathrm{l}}$ and $M_{\mathrm{u}}$ for the lower and upper mass limits,
respectively). The volumetric rate of CC SNe, $R_{\mathrm{CC}}$ in units of 
yr$^{-1}$ Mpc$^{-3}$ is then:
\begin{equation}
R_{\mathrm{CC}}(z)=  k\times \rho_*(z),
\label{eq:snr} 
\end{equation}
where $\rho_*(z)$ is the star formation history in units of
$M_{\sun}$ yr$^{-1}$ Mpc$^{-3}$. $R_{\mathrm{CC}}$ is a volumetric
rate, and thus scales as h$^3$, this needs to be addressed when comparing
to star formation histories which may have a different scaling. The constant
$k$ is the percentage of stars that explode as SNe per unit mass and is given by:
\begin{equation}
k = \frac{\int^{M_{\mathrm{u}}}_{M_{\mathrm{l}}} \xi(M) 
dM}{\int^{125 M_{\sun}}_{0.1 M_{\sun}} M \xi(M) dM}.
\end{equation}
For this work we choose to use a Salpeter IMF
\citep{1955ApJ...121..161S} with progenitor masses between
$8 M_{\sun}$ and $50 M_{\sun}$. The constant $k$ is
then equal to 0.0070.

The choice to use a Salpeter IMF is based on that most SFH measurements
and other supernova rate measurements are using this IMF. Recently a
number of authors have argued that the Salpeter IMF is not consistent
with recent observations \citep[see, e.g.][]{2006ApJ...651..142H}, and
claim that an IMF with a flatter shape is better. The choice to use
different IMFs introduce a systematic uncertainty to the comparison.
However, the IMF dependence is partly cancelled out by the dependence on
IMF of the star formation history (where the IMF is used to normalise
the star formation measurements). To investigate this we compute $k$ for
the flattest IMF given in \citet{2006ApJ...651..142H} which is a
modified Salpeter form from \citet{2003ApJ...593..258B} with a high mass
slope of $-1.15$ (compared to $-1.35$ for the normal Salpeter IMF). While
the $k$ factor changes to 0.0141 (a factor of $\sim 2$) this is
countered by the change of the SFH normalisation by a factor of 0.5,
given by \citet{2006ApJ...651..142H}. The total change of the SNR--SFR
comparison is on the order of 2\%, significantly less than the
statistical errors. Incidentally, this also tells us that there is
little hope in trying to use the comparisons between SNR and star
formation rates to constrain a possible universal IMF unless much
larger supernova samples can be obtained.

The upper limit for the progenitor mass comes from
\citet{1997ApJ...483..228T} and is essentially the limit at which
massive stars will produce black holes instead of neutron stars with an
associated supernova explosion.  Varying the upper limit between
$M_{\mathrm{u}}=\left[30,125\right] M_{\sun}$ has a small
effect on the SNR--SFR comparison, less than 10\%, although this could
be slightly higher with a flatter IMF. The lower limit of
$8\, M_{\sun}$ is grounded on observations of supernova
progenitors \citep{2009MNRAS.395.1409S}. The choice of lower limit
strongly affects the rate scaling but is luckily well constrained by
these observations, varying the lower limit between their 68\%
confidence limits (7--9.5 $M_{\sun}$) changes $k$ by
$\pm$20\%, also this lower than the statistical errors for our supernova
rates.
\subsubsection{Star formation history parametrisations}
\label{sfh}
We use star formation histories, $\rho_*(z)$, from two different sources
as input for Eq.~\ref{eq:snr}. \citet{2006ApJ...651..142H} compile
star formation measurements from many sources and correct them for dust
extinction based on the work done by \citet{2005ApJ...632..169L}. The
observations are then fit by the simple analytical parametrisation
introduced by \citet{2001MNRAS.326..255C}:
\begin{equation}
\rho_*(z) = \frac{(a+bz)\,\mbox{h}_{70}}{1+(z/c)^d},
\label{eq:hbsfh}
\end{equation}
where $a$, $b$, $c$ and $d$ are fitting parameters and h$_{70}$ is the Hubble
constant to account for changes to the assumed cosmology. The best fit
from \citet{2006ApJ...651..142H} gives parameter values: $a=0.0170$,
$b=0.13$, $c=3.3$ and $d=5.3$. They give a conservative error estimate
by plotting the confidence regions of the fit, at $z\sim 1$ the
uncertainty is on the order of 30\%.

\citet{2004ApJ...613..200S} fit the data compilation of 
\citet{2004ApJ...600L.103G} by using a parametrisation of the form:
\begin{equation}
\rho_*(t) = a(t^b e^{-t/c} + d e^{d(t-t_0)/c}),
\label{eq:giasfh}
\end{equation}
where $a$, $b$, $c$ and $d$ are parameters, $t$ is the age of 
the universe and $t_0$ the age of the universe at $z=0$ (13.47 Gyrs 
with our chosen cosmology). The best fit parameter values are: $a=0.182$, 
$b=1.26$, $c=1.865$ and $d=0.071$. To compare this with our results
we convert it to $\rho_*(z)$.
\subsection{Delay time distributions for Ia supernovae}
\label{sec:dtd}
The delay time distribution (DTD) for Ia supernovae is the subject of 
controversy in current research. By
convolving an assumed DTD with the underlying star formation history the
Ia rate can be calculated. This can then be compared with the observed
rates to put constraints on the DTDs. The Ia rate can thus be written 
\citep{2004ApJ...613..200S}:
\begin{equation}
R_{\mathrm{Ia}}(t) = \nu \int^t_{t_0} \rho_*(t') \Phi(t-t') dt',
\end{equation}
where $\nu$ is the number of SNe formed per unit stellar mass,
$\rho_*(t)$ the star formation history and $\Phi(\tau)$ the delay
time distribution which represents the percentage of supernovae that go
off at time $\tau$ after a single burst of star formation. Also,
$t_0$ is the time at which stars start to form in the universe, we
assume this happens at $z=10$; corresponding to a time of 0.45 Gyrs
after the big bang (with our chosen cosmology). The integral can easily
be converted to redshift space to yield $R_{\mathrm{Ia}}(z)$ by performing a 
variable substitution.

As noted above, the uncertainties in our rate estimates make it quite
fruitless to try and fit different DTDs to our data. Instead we choose
to compare our rates to the best-fitting DTDs from the literature. In
Fig.~\ref{fig:Iarates} we show four different choices of DTD. The
simplest one is just a rescaling of the star formation history assuming
that one Ia SN explode per $1000\, M_{\sun}$ of stellar mass formed
\citep{2006AJ....132.1126N}. This corresponds to using a Dirac delta 
function DTD with the peak at $\tau=0$, i.e. no delay time between
the SN explosion and star formation or, in other words, a prompt channel 
for the Ia progenitor-to-supernova process.

In \citet{2010ApJ...713...32S} and \citet{2004ApJ...613..200S} a
Gaussian shaped (or close to Gaussian) DTD with a mean delay time of
$\sim 3$ Gyrs is found to give the best fit to the GOODs Ia supernova
rates. We use the unimodal skew-normal DTD parametrisation and best
fitting parameters from \citet{2010ApJ...713...32S} and convolve this
with the star formation history given in Eq.~\ref{eq:giasfh} (the
same SFH as the one used by these authors). The DTD parametrisation is:
\begin{equation}
\Phi(\tau) = \frac{1}{\omega \pi} e^{\frac{-(\tau - \xi)^2}{2\omega^2}}
\int_{-\inf}^{\alpha \left(\frac{\tau-\xi}{\omega}\right)} 
e^{-t'^2/2} \, dt',
\end{equation}
with the best fitting parameters $\omega=0.2$, $\xi=3.2$ and 
$\alpha=2.2$.

\citet{2011MNRAS.tmp.1508G} fit the results from the Subaru Deep Field supernova
search together with the rates of many other surveys using different
power law DTDs.  The DTD models and SFHs they try show quite similar
fitting quality, the rate we show in Fig.~\ref{fig:Iarates} uses a
power law DTD with an exponent $\beta$ equal to 0.97, their best fitting
value:
\begin{equation}
\Phi(\tau)= \Phi_1 \tau^{\beta},
\end{equation}
where $\Phi_1$ is a normalisation parameter. $\Phi$ is also set to 0 for
$\tau<40$ million years. This is then convolved with the SFH given in
Eq.~\ref{eq:hbsfh} to get the rates 
plotted in the Fig.~\ref{fig:Iarates}.

\citet{2005ApJ...629L..85S}, \citet{2006MNRAS.370..773M} and
\citet{2006ApJ...648..868S} find strong evidence for the existence of a
prompt channel for Ia SNe by looking at the properties of the host
galaxies of Ia SNe (and other data). \citet{2011MNRAS.tmp..307M} study
the hosts of the local SNe found in the LOSS and find a prompt channel
(with $\tau<420$ Myrs) with 99\% significance.  Models of Ia explosions
also show the possibility of such a channel
\citep{2009AIPC.1111..267N,2011MNRAS.417..408R}. The Ia rates are then
modelled with one prompt component, directly proportional to the SFH,
and one delayed component. We plot the resulting Ia rate from the work
by \citet{2006ApJ...648..868S}, who use a parametrisation of the form
\begin{equation}
R_{\mathrm{Ia}}(t)= A \int_0^t \tilde{\rho}_*(t')\, dt' + B\tilde{\rho}_*(t).
\end{equation}
Note that the SFH used by us is not strictly the same as
the one used by these authors and can thus not be used 
directly in this equation. We plot the SNR as
given in \citet{2006ApJ...648..868S} in Fig.~\ref{fig:Iarates}.
\section{Discussion and summary}
\label{sec:disc}
We have presented supernova rates from the SVISS along with a description of
the methods used to compute them and an extensive analysis of systematic errors
for the rates. The resulting rates for the core collapse SNe are:
$3.29_{-1.78\,-1.45}^{+3.08\,+1.98}\times 10^{-4}$ yr$^{-1}$ Mpc$^{-3}$
h$^3_{70}$ (with statistical and systematic errors respectively) at $\langle
z\rangle =0.39$ and $6.40_{-3.12\,-2.11}^{+5.30\,+3.65}\times 10^{-4}$
yr$^{-1}$ Mpc$^{-3}$ h$^3_{70}$ at $\langle z\rangle=0.73$. The CC rates have
been corrected for obscuration in dusty environments using the results of
\citet{2012arXiv1206.1314M}, as described in Sect.~\ref{sec:obsc} and host
galaxy extinction based on the models in \citet{2005MNRAS.362..671R}.
Uncorrected values can be found in Table~\ref{table:rates}.

The rate estimates follow the star formation history well.
\citet{2011ApJ...738..154H} point out that the core collapse supernova
rates found in both local and high redshift searches seem to be too low
by a factor of two when compared to the star formation history compiled
in \citet{2006ApJ...651..142H}. Our results do not show this difference,
both of our rate estimates are consistent with the star formation
history within the statistical errors. But our errors are 
quite large, the rates estimated by other authors, in particular at low
$z$, do in fact differ significantly from the SFH. The strongest
evidence for this comes from the rate estimate by
\citet{2011MNRAS.tmp..317L} at low $z$ which has small errors and is
clearly below the SFH. At higher redshift the problem is less severe,
which could be due to the increased statistical errors. At
high redshift the problem may also be somewhat alleviated by the
obscuration corrections, which we have included in our plotted rates in
Fig.~\ref{fig:ccrates}, the rates from other surveys plotted in this
figure do not include this correction (with the exception of the data
point from \citealt{2011MNRAS.tmp.1508G}). 

\citet{2011ApJ...738..154H} suggest that taking missed SNe due to extinction
and dust obscuration in LIRGs and ULIRGs into account is not enough to explain
the difference. Instead they suggest that the reason is that the assumed
fractions of faint and very faint CC SNe are too low. In our tests of the
systematic uncertainties we find that assuming 30\% of the CC SNe to be faint
($M_B>-15$) boosts the rates by $\sim$30\%, not enough to bridge the factor of
two gap found for the local SN searches.  Of course, the assumption on
fractions of faint CC SNe may be very different for the different surveys,
making it possible that other data points may go up more than this. While the
SFHs we compare with have also been corrected for dust and hidden star
formation, the supernovae are probably affected differently by the presence of
obscuring dust.  The number of supernovae that are missed in dusty starburst
galaxies and LIRGs is currently not well constrained even in the local universe
\citep[e.g][]{2003A&A...401..519M, 2004NewAR..48..595M}. The de-bias factors
for extinction and dust obscuration in normal galaxies and U/LIRGs derived by
\citet{2012arXiv1206.1314M} can make the difference between the predicted and
observed CC SN rates disappear.  In this paper we have used these factors to
de-bias the core collapse supernova rates. The de-bias factors from this study
are slightly larger at low redshifts, but lower at high redshift, than the
factors from \citet{2007MNRAS.377.1229M}.  The uncertainties -- both
statistical and systematic -- of the de-bias factors derived in
\citet{2012arXiv1206.1314M} have been thoroughly studied, and will hopefully
decrease as more observations of SNe in dusty galaxies are obtained. High
angular resolution observations at near-IR \citep[e.g.][]{2007ApJ...659L...9M,
2008ApJ...689L..97K, 2012ApJ...744L..19K} and radio
\citep[e.g,][]{2010A&A...519L...5P, 2011MNRAS.415.2688R} wavelengths have
recently been used to detect and characterise the hidden SN populations in the
nearest LIRGs. Such studies are needed to constrain the complete rates,
properties and extinction distributions towards the CC SNe buried in such dusty
galaxies. Eventually, these studies will hopefully provide a robust estimate
for the numbers of SNe missed by optical searches both locally and at high-z. 

The resulting rate for the Ia supernovae is:
1.29$_{-0.57\,-0.28}^{+0.88\,+0.27}\times 10^{-4}$ yr$^{-1}$ Mpc$^{-3}$
h$^3_{70}$ at $\langle z\rangle =0.62$. This rate has been corrected for host
galaxy extinction, but not for high redshift obscuration (a positive correction
for this is included in the systematic error).

Because of the quite high statistical errors and the lack of Ia SNe beyond
$z=1$ we do not try to fit any DTDs to our Ia rate measurements. The comparison
to the plotted models show that all these models are consistent with our rates.
The high rate of Ia SNe at $z\sim 0.5$ measured by SVISS is in strong agreement
with the results of \citet{2008ApJ...681..462D}, and thus in support of a
Gaussian-like, fairly wide, DTD for Ia SNe. However, it should be noted that
our measurement is also consistent with the distributions proposed by other
authors.  The type Ia SN rate measurements at $z\gtrsim 0.4$ differ by up to a
factor of $\sim 2$. We believe that the cause of this large scatter is the
large statistical and systematic errors that the SN surveys suffer from in this
redshift range. It should also be noted that there are differences in the
methods used to calculate the rates in the different surveys. It is therefore
of utmost importance that the systematic errors are correctly estimated.
\citet{2010ApJ...713...32S} find that models with a prompt Ia component are
hard to reconcile with the rates measured at redshifts higher than one.
Increased sample sizes at these high redshifts, or more studies of Ia host
galaxies \citep{2005ApJ...634..210G,2006ApJ...648..868S} are needed to
constrain the contribution from this channel.

In Section~\ref{sec:debiasIa} we outline our motivation for not de-biasing the
type Ia SN rates to account for extremely high extinction in U/LIRGs. It is
very important to understand that the missing fractions of type Ia SNe situated
in U/LIRGs similar to Arp~299 very likely depend on the DTDs.  It is likely
that also the missing fractions in other types of galaxies will depend on the
DTD, although to a lesser extent. Using the type Ia SN rates to estimate the
DTD thus inherently suffers from circularity to some extent.  Assuming a zero
missing fraction is a choice in itself -- the DTD implicit to such an
assumption has a cut-off at some delay time $\tau \lesssim 200$ Myrs.  If the
DTD has no such cut-off, and a significant number of type Ia SNe have very
short delay times, our assumption is faulty. In that case the resulting rates
may be off by more than the adopted systematic error.

The determination of supernova rates at high redshift is difficult. The
SNe detected at high redshift will only be sampling the bright end of the SN
luminosity function. This means that any global statistics estimated from such
measurements will be sensitive to assumptions on the luminosity function
made during the calculations. There are a number of additional assumptions
going into the rate calculations that affect the rates differently. It is
important to estimate the systematic errors these assumptions give rise to.
With the exception of the misclassification error, that essentially scales with
the sample size for samples smaller than ten given a misclassification ratio of
10\%, we have shown that the systematic errors are on the order of 50\% when
using photometric redshifts and with the present uncertainties in template
fractions and peak magnitudes. Furthermore, the assumptions made when
correcting the rates for extinction/obscuration are shown to have a
large effect on the systematics. Presently little is known about the number of
SNe missed in LIRGs and ULIRGs, especially important at high redshift,
which is evident in the large uncertainties on the de-bias factors.

Given the low numbers of SNe for most high redshift surveys it is
perhaps tempting to try and use the rates found in different surveys
together when comparing to models (for the Ia SNe) and other sources
(SFH for CC SNe). However, it's not straight-forward to do this.
Different surveys estimate systematic errors in different ways, some include more
sources and some less.  If combined as given using co-added statistical
and systematic errors, the risk is that greater weight is given to
surveys where fewer systematic error estimates are included (given that
the sample sizes are similar). We believe that the work presented in
this paper shows the importance of including a variety of systematic
effects to correctly estimate the uncertainties of supernova rates at
high redshift. This will be even more important for future surveys with 
larger sample sizes and therefore lower statistical errors.

\begin{acknowledgements}
We are grateful for financial support from the Swedish Research 
Council. S.~M., J.~M., and E.~K. acknowledge financial support from the Academy 
of Finland (project:8120503). We also thank D.~F. de Mello for interesting
input on the manuscript. We thank the anonymous referee for 
useful comments.  

\end{acknowledgements}
\bibliographystyle{aa}
\bibliography{database}

\begin{thebibliography}{99}
\expandafter\ifx\csname natexlab\endcsname\relax\def\natexlab#1{#1}\fi

\bibitem[{{Alard}(2000)}]{2000A&AS..144..363A}
{Alard}, C. 2000, \aaps, 144, 363, (A00)

\bibitem[{{Astier} {et~al.}(2006){Astier}, {Guy}, {Regnault}, {Pain},
  {Aubourg}, {Balam}, {Basa}, {Carlberg}, {Fabbro}, {Fouchez}, {Hook},
  {Howell}, {Lafoux}, {Neill}, {Palanque-Delabrouille}, {Perrett}, {Pritchet},
  {Rich}, {Sullivan}, {Taillet}, {Aldering}, {Antilogus}, {Arsenijevic},
  {Balland}, {Baumont}, {Bronder}, {Courtois}, {Ellis}, {Filiol}, {Gon{\c
  c}alves}, {Goobar}, {Guide}, {Hardin}, {Lusset}, {Lidman}, {McMahon},
  {Mouchet}, {Mourao}, {Perlmutter}, {Ripoche}, {Tao}, \&
  {Walton}}]{2006A&A...447...31A}
{Astier}, P., {Guy}, J., {Regnault}, N., {et~al.} 2006, \aap, 447, 31

\bibitem[{{Baldry} \& {Glazebrook}(2003)}]{2003ApJ...593..258B}
{Baldry}, I.~K. \& {Glazebrook}, K. 2003, \apj, 593, 258

\bibitem[{{Barbary} {et~al.}(2012{\natexlab{a}}){Barbary}, {Aldering},
  {Amanullah}, {Brodwin}, {Connolly}, {Dawson}, {Doi}, {Eisenhardt},
  {Faccioli}, {Fadeyev}, {Fakhouri}, {Fruchter}, {Gilbank}, {Gladders},
  {Goldhaber}, {Goobar}, {Hattori}, {Hsiao}, {Huang}, {Ihara}, {Kashikawa},
  {Koester}, {Konishi}, {Kowalski}, {Lidman}, {Lubin}, {Meyers}, {Morokuma},
  {Oda}, {Panagia}, {Perlmutter}, {Postman}, {Ripoche}, {Rosati}, {Rubin},
  {Schlegel}, {Spadafora}, {Stanford}, {Strovink}, {Suzuki}, {Takanashi},
  {Tokita}, {Yasuda}, \& {Supernova Cosmology Project}}]{2012ApJ...745...32B}
{Barbary}, K., {Aldering}, G., {Amanullah}, R., {et~al.} 2012{\natexlab{a}},
  \apj, 745, 32

\bibitem[{{Barbary} {et~al.}(2012{\natexlab{b}}){Barbary}, {Aldering},
  {Amanullah}, {Brodwin}, {Connolly}, {Dawson}, {Doi}, {Eisenhardt},
  {Faccioli}, {Fadeyev}, {Fakhouri}, {Fruchter}, {Gilbank}, {Gladders},
  {Goldhaber}, {Goobar}, {Hattori}, {Hsiao}, {Huang}, {Ihara}, {Kashikawa},
  {Koester}, {Konishi}, {Kowalski}, {Lidman}, {Lubin}, {Meyers}, {Morokuma},
  {Oda}, {Panagia}, {Perlmutter}, {Postman}, {Ripoche}, {Rosati}, {Rubin},
  {Schlegel}, {Spadafora}, {Stanford}, {Strovink}, {Suzuki}, {Takanashi},
  {Tokita}, {Yasuda}, \& {Supernova Cosmology Project}}]{2012ApJ...745...31B}
{Barbary}, K., {Aldering}, G., {Amanullah}, R., {et~al.} 2012{\natexlab{b}},
  \apj, 745, 31

\bibitem[{{Bazin} {et~al.}(2009){Bazin}, {Palanque-Delabrouille}, {Rich},
  {Ruhlmann-Kleider}, {Aubourg}, {Le Guillou}, {Astier}, {Balland}, {Basa},
  {Carlberg}, {Conley}, {Fouchez}, {Guy}, {Hardin}, {Hook}, {Howell}, {Pain},
  {Perrett}, {Pritchet}, {Regnault}, {Sullivan}, {Antilogus}, {Arsenijevic},
  {Baumont}, {Fabbro}, {Le Du}, {Lidman}, {Mouchet}, {Mour{\~a}o}, \&
  {Walker}}]{2009A&A...499..653B}
{Bazin}, G., {Palanque-Delabrouille}, N., {Rich}, J., {et~al.} 2009, \aap, 499,
  653

\bibitem[{{Bertin} \& {Arnouts}(1996)}]{1996A&AS..117..393B}
{Bertin}, E. \& {Arnouts}, S. 1996, \aaps, 117, 393

\bibitem[{{Botticella} {et~al.}(2008){Botticella}, {Riello}, {Cappellaro},
  {Benetti}, {Altavilla}, {Pastorello}, {Turatto}, {Greggio}, {Patat},
  {Valenti}, {Zampieri}, {Harutyunyan}, {Pignata}, \&
  {Taubenberger}}]{2008A&A...479...49B}
{Botticella}, M.~T., {Riello}, M., {Cappellaro}, E., {et~al.} 2008, \aap, 479,
  49

\bibitem[{{Botticella} {et~al.}(2012){Botticella}, {Smartt}, {Kennicutt},
  {Cappellaro}, {Sereno}, \& {Lee}}]{2012A&A...537A.132B}
{Botticella}, M.~T., {Smartt}, S.~J., {Kennicutt}, R.~C., {et~al.} 2012, \aap,
  537, A132

\bibitem[{{Bouwens} {et~al.}(2009){Bouwens}, {Illingworth}, {Franx}, {Chary},
  {Meurer}, {Conselice}, {Ford}, {Giavalisco}, \& {van
  Dokkum}}]{2009ApJ...705..936B}
{Bouwens}, R.~J., {Illingworth}, G.~D., {Franx}, M., {et~al.} 2009, \apj, 705,
  936

\bibitem[{{Calzetti} {et~al.}(2000){Calzetti}, {Armus}, {Bohlin}, {Kinney},
  {Koornneef}, \& {Storchi-Bergmann}}]{2000ApJ...533..682C}
{Calzetti}, D., {Armus}, L., {Bohlin}, R.~C., {et~al.} 2000, \apj, 533, 682

\bibitem[{{Cappellaro} {et~al.}(1999){Cappellaro}, {Evans}, \&
  {Turatto}}]{1999A&A...351..459C}
{Cappellaro}, E., {Evans}, R., \& {Turatto}, M. 1999, \aap, 351, 459

\bibitem[{{Cappellaro} {et~al.}(2005){Cappellaro}, {Riello}, {Altavilla},
  {Botticella}, {Benetti}, {Clocchiatti}, {Danziger}, {Mazzali}, {Pastorello},
  {Patat}, {Salvo}, {Turatto}, \& {Valenti}}]{2005A&A...430...83C}
{Cappellaro}, E., {Riello}, M., {Altavilla}, G., {et~al.} 2005, \aap, 430, 83

\bibitem[{{Cardelli} {et~al.}(1989){Cardelli}, {Clayton}, \&
  {Mathis}}]{1989ApJ...345..245C}
{Cardelli}, J.~A., {Clayton}, G.~C., \& {Mathis}, J.~S. 1989, \apj, 345, 245

\bibitem[{{Cole} {et~al.}(2001){Cole}, {Norberg}, {Baugh}, {Frenk},
  {Bland-Hawthorn}, {Bridges}, {Cannon}, {Colless}, {Collins}, {Couch},
  {Cross}, {Dalton}, {De Propris}, {Driver}, {Efstathiou}, {Ellis},
  {Glazebrook}, {Jackson}, {Lahav}, {Lewis}, {Lumsden}, {Maddox}, {Madgwick},
  {Peacock}, {Peterson}, {Sutherland}, \& {Taylor}}]{2001MNRAS.326..255C}
{Cole}, S., {Norberg}, P., {Baugh}, C.~M., {et~al.} 2001, \mnras, 326, 255

\bibitem[{{Coleman} {et~al.}(1980){Coleman}, {Wu}, \&
  {Weedman}}]{1980ApJS...43..393C}
{Coleman}, G.~D., {Wu}, C., \& {Weedman}, D.~W. 1980, \apjs, 43, 393

\bibitem[{{Daddi} {et~al.}(2010){Daddi}, {Bournaud}, {Walter}, {Dannerbauer},
  {Carilli}, {Dickinson}, {Elbaz}, {Morrison}, {Riechers}, {Onodera}, {Salmi},
  {Krips}, \& {Stern}}]{2010ApJ...713..686D}
{Daddi}, E., {Bournaud}, F., {Walter}, F., {et~al.} 2010, \apj, 713, 686

\bibitem[{{Dahl{\'e}n} \& {Fransson}(1999)}]{1999A&A...350..349D}
{Dahl{\'e}n}, T. \& {Fransson}, C. 1999, \aap, 350, 349

\bibitem[{{Dahlen} {et~al.}(2010){Dahlen}, {Mobasher}, {Dickinson}, {Ferguson},
  {Giavalisco}, {Grogin}, {Guo}, {Koekemoer}, {Lee}, {Lee}, {Nonino}, {Riess},
  \& {Salimbeni}}]{2010ApJ...724..425D}
{Dahlen}, T., {Mobasher}, B., {Dickinson}, M., {et~al.} 2010, \apj, 724, 425

\bibitem[{{Dahlen} {et~al.}(2008){Dahlen}, {Strolger}, \&
  {Riess}}]{2008ApJ...681..462D}
{Dahlen}, T., {Strolger}, L.-G., \& {Riess}, A.~G. 2008, \apj, 681, 462

\bibitem[{{Dahlen} {et~al.}(2004){Dahlen}, {Strolger}, {Riess}, {Mobasher},
  {Chary}, {Conselice}, {Ferguson}, {Fruchter}, {Giavalisco}, {Livio}, {Madau},
  {Panagia}, \& {Tonry}}]{2004ApJ...613..189D}
{Dahlen}, T., {Strolger}, L.-G., {Riess}, A.~G., {et~al.} 2004, \apj, 613, 189

\bibitem[{{Dilday} {et~al.}(2010{\natexlab{a}}){Dilday}, {Bassett}, {Becker},
  {Bender}, {Castander}, {Cinabro}, {Frieman}, {Galbany}, {Garnavich},
  {Goobar}, {Hopp}, {Ihara}, {Jha}, {Kessler}, {Lampeitl}, {Marriner},
  {Miquel}, {Moll{\'a}}, {Nichol}, {Nordin}, {Riess}, {Sako}, {Schneider},
  {Smith}, {Sollerman}, {Wheeler}, {{\"O}stman}, {Bizyaev}, {Brewington},
  {Malanushenko}, {Malanushenko}, {Oravetz}, {Pan}, {Simmons}, \&
  {Snedden}}]{2010ApJ...715.1021D}
{Dilday}, B., {Bassett}, B., {Becker}, A., {et~al.} 2010{\natexlab{a}}, \apj,
  715, 1021

\bibitem[{{Dilday} {et~al.}(2010{\natexlab{b}}){Dilday}, {Smith}, {Bassett},
  {Becker}, {Bender}, {Castander}, {Cinabro}, {Filippenko}, {Frieman},
  {Galbany}, {Garnavich}, {Goobar}, {Hopp}, {Ihara}, {Jha}, {Kessler},
  {Lampeitl}, {Marriner}, {Miquel}, {Moll{\'a}}, {Nichol}, {Nordin}, {Riess},
  {Sako}, {Schneider}, {Sollerman}, {Wheeler}, {{\"O}stman}, {Bizyaev},
  {Brewington}, {Malanushenko}, {Malanushenko}, {Oravetz}, {Pan}, {Simmons}, \&
  {Snedden}}]{2010ApJ...713.1026D}
{Dilday}, B., {Smith}, M., {Bassett}, B., {et~al.} 2010{\natexlab{b}}, \apj,
  713, 1026

\bibitem[{{Elbaz} {et~al.}(2011){Elbaz}, {Dickinson}, {Hwang},
  {D{\'{\i}}az-Santos}, {Magdis}, {Magnelli}, {Le Borgne}, {Galliano},
  {Pannella}, {Chanial}, {Armus}, {Charmandaris}, {Daddi}, {Aussel}, {Popesso},
  {Kartaltepe}, {Altieri}, {Valtchanov}, {Coia}, {Dannerbauer}, {Dasyra},
  {Leiton}, {Mazzarella}, {Alexander}, {Buat}, {Burgarella}, {Chary}, {Gilli},
  {Ivison}, {Juneau}, {Le Floc'h}, {Lutz}, {Morrison}, {Mullaney}, {Murphy},
  {Pope}, {Scott}, {Brodwin}, {Calzetti}, {Cesarsky}, {Charlot}, {Dole},
  {Eisenhardt}, {Ferguson}, {F{\"o}rster Schreiber}, {Frayer}, {Giavalisco},
  {Huynh}, {Koekemoer}, {Papovich}, {Reddy}, {Surace}, {Teplitz}, {Yun}, \&
  {Wilson}}]{2011A&A...533A.119E}
{Elbaz}, D., {Dickinson}, M., {Hwang}, H.~S., {et~al.} 2011, \aap, 533, A119

\bibitem[{{Elias-Rosa} {et~al.}(2011){Elias-Rosa}, {Van Dyk}, {Li},
  {Silverman}, {Foley}, {Ganeshalingam}, {Mauerhan}, {Kankare}, {Jha},
  {Filippenko}, {Beckman}, {Berger}, {Cuillandre}, \&
  {Smith}}]{2011ApJ...742....6E}
{Elias-Rosa}, N., {Van Dyk}, S.~D., {Li}, W., {et~al.} 2011, \apj, 742, 6

\bibitem[{{Gallagher} {et~al.}(2005){Gallagher}, {Garnavich}, {Berlind},
  {Challis}, {Jha}, \& {Kirshner}}]{2005ApJ...634..210G}
{Gallagher}, J.~S., {Garnavich}, P.~M., {Berlind}, P., {et~al.} 2005, \apj,
  634, 210

\bibitem[{{Gehrels}(1986)}]{1986ApJ...303..336G}
{Gehrels}, N. 1986, \apj, 303, 336

\bibitem[{{Giavalisco} {et~al.}(2004){Giavalisco}, {Dickinson}, {Ferguson},
  {Ravindranath}, {Kretchmer}, {Moustakas}, {Madau}, {Fall}, {Gardner},
  {Livio}, {Papovich}, {Renzini}, {Spinrad}, {Stern}, \&
  {Riess}}]{2004ApJ...600L.103G}
{Giavalisco}, M., {Dickinson}, M., {Ferguson}, H.~C., {et~al.} 2004, \apjl,
  600, L103

\bibitem[{{Goobar}(2008)}]{2008ApJ...686L.103G}
{Goobar}, A. 2008, \apjl, 686, L103

\bibitem[{{Goto}(2007)}]{2007MNRAS.377.1222G}
{Goto}, T. 2007, \mnras, 377, 1222

\bibitem[{{Graur} {et~al.}(2011){Graur}, {Poznanski}, {Maoz}, {Yasuda},
  {Totani}, {Fukugita}, {Filippenko}, {Foley}, {Silverman}, {Gal-Yam},
  {Horesh}, \& {Jannuzi}}]{2011MNRAS.tmp.1508G}
{Graur}, O., {Poznanski}, D., {Maoz}, D., {et~al.} 2011, \mnras, 1508

\bibitem[{{Greggio}(2005)}]{2005A&A...441.1055G}
{Greggio}, L. 2005, \aap, 441, 1055

\bibitem[{{Hatano} {et~al.}(1998){Hatano}, {Branch}, \&
  {Deaton}}]{1998ApJ...502..177H}
{Hatano}, K., {Branch}, D., \& {Deaton}, J. 1998, \apj, 502, 177

\bibitem[{{Hayes} {et~al.}(2010){Hayes}, {Schaerer}, \&
  {{\"O}stlin}}]{2010A&A...509L...5H}
{Hayes}, M., {Schaerer}, D., \& {{\"O}stlin}, G. 2010, \aap, 509, L5

\bibitem[{{Hopkins} \& {Beacom}(2006)}]{2006ApJ...651..142H}
{Hopkins}, A.~M. \& {Beacom}, J.~F. 2006, \apj, 651, 142

\bibitem[{{Horiuchi} {et~al.}(2011){Horiuchi}, {Beacom}, {Kochanek}, {Prieto},
  {Stanek}, \& {Thompson}}]{2011ApJ...738..154H}
{Horiuchi}, S., {Beacom}, J.~F., {Kochanek}, C.~S., {et~al.} 2011, \apj, 738,
  154

\bibitem[{{Kankare} {et~al.}(2008){Kankare}, {Mattila}, {Ryder},
  {P{\'e}rez-Torres}, {Alberdi}, {Romero-Canizales}, {D{\'{\i}}az-Santos},
  {V{\"a}is{\"a}nen}, {Efstathiou}, {Alonso-Herrero}, {Colina}, \&
  {Kotilainen}}]{2008ApJ...689L..97K}
{Kankare}, E., {Mattila}, S., {Ryder}, S., {et~al.} 2008, \apjl, 689, L97

\bibitem[{{Kankare} {et~al.}(2012){Kankare}, {Mattila}, {Ryder},
  {V{\"a}is{\"a}nen}, {Alberdi}, {Alonso-Herrero}, {Colina}, {Efstathiou},
  {Kotilainen}, {Melinder}, {P{\'e}rez-Torres}, {Romero-Ca{\~n}izales}, \&
  {Takalo}}]{2012ApJ...744L..19K}
{Kankare}, E., {Mattila}, S., {Ryder}, S., {et~al.} 2012, \apjl, 744, L19

\bibitem[{{Kartaltepe} {et~al.}(2011){Kartaltepe}, {Dickinson}, {Alexander},
  {Bell}, {Dahlen}, {Elbaz}, {Faber}, {Lotz}, {McIntosh}, {Wiklind}, {Altieri},
  {Aussel}, {Bethermin}, {Bournaud}, {Charmandaris}, {Conselice}, {Cooray},
  {Daddi}, {Dannerbauer}, {Dav{\'e}}, {Dunlop}, {Dekel}, {Ferguson}, {Grogin},
  {Hwang}, {Ivison}, {Kocevski}, {Koekemoer}, {Koo}, {Lai}, {Leiton}, {Lucas},
  {Lutz}, {Magdis}, {Magnelli}, {Morrison}, {Mozena}, {Mullaney}, {Newman},
  {Pope}, {Popesso}, {van der Wel}, {Weiner}, \& {Wuyts}}]{2011arXiv1110.4057K}
{Kartaltepe}, J.~S., {Dickinson}, M., {Alexander}, D.~M., {et~al.} 2011,
  [arXiv:1110.4057]

\bibitem[{{Kessler} {et~al.}(2010){Kessler}, {Bassett}, {Belov}, {Bhatnagar},
  {Campbell}, {Conley}, {Frieman}, {Glazov}, {Gonz{\'a}lez-Gait{\'a}n},
  {Hlozek}, {Jha}, {Kuhlmann}, {Kunz}, {Lampeitl}, {Mahabal}, {Newling},
  {Nichol}, {Parkinson}, {Philip}, {Poznanski}, {Richards}, {Rodney}, {Sako},
  {Schneider}, {Smith}, {Stritzinger}, \& {Varughese}}]{2010PASP..122.1415K}
{Kessler}, R., {Bassett}, B., {Belov}, P., {et~al.} 2010, \pasp, 122, 1415

\bibitem[{{Kinney} {et~al.}(1996){Kinney}, {Calzetti}, {Bohlin}, {McQuade},
  {Storchi-Bergmann}, \& {Schmitt}}]{1996ApJ...467...38K}
{Kinney}, A.~L., {Calzetti}, D., {Bohlin}, R.~C., {et~al.} 1996, \apj, 467, 38

\bibitem[{{Kuznetsova} \& {Connolly}(2007)}]{2007ApJ...659..530K}
{Kuznetsova}, N.~V. \& {Connolly}, B.~M. 2007, \apj, 659, 530

\bibitem[{{La Franca} {et~al.}(2004){La Franca}, {Gruppioni}, {Matute},
  {Pozzi}, {Lari}, {Mignoli}, {Zamorani}, {Alexander}, {Cocchia}, {Danese},
  {Franceschini}, {H{\'e}raudeau}, {Kotilainen}, {Linden-V{\o}rnle}, {Oliver},
  {Rowan-Robinson}, {Serjeant}, {Spinoglio}, \& {Verma}}]{2004AJ....127.3075L}
{La Franca}, F., {Gruppioni}, C., {Matute}, I., {et~al.} 2004, \aj, 127, 3075

\bibitem[{{Le F{\`e}vre} {et~al.}(2003){Le F{\`e}vre}, {Saisse}, {Mancini},
  {Brau-Nogue}, {Caputi}, {Castinel}, {D'Odorico}, {Garilli}, {Kissler-Patig},
  {Lucuix}, {Mancini}, {Pauget}, {Sciarretta}, {Scodeggio}, {Tresse}, \&
  {Vettolani}}]{2003SPIE.4841.1670L}
{Le F{\`e}vre}, O., {Saisse}, M., {Mancini}, D., {et~al.} 2003, in \procspie,
  Vol. 4841, 1670

\bibitem[{{Le Floc'h} {et~al.}(2005){Le Floc'h}, {Papovich}, {Dole}, {Bell},
  {Lagache}, {Rieke}, {Egami}, {P{\'e}rez-Gonz{\'a}lez}, {Alonso-Herrero},
  {Rieke}, {Blaylock}, {Engelbracht}, {Gordon}, {Hines}, {Misselt}, {Morrison},
  \& {Mould}}]{2005ApJ...632..169L}
{Le Floc'h}, E., {Papovich}, C., {Dole}, H., {et~al.} 2005, \apj, 632, 169

\bibitem[{{Leibundgut}(2000)}]{2000A&ARv..10..179L}
{Leibundgut}, B. 2000, \aapr, 10, 179

\bibitem[{{Li} {et~al.}(2011{\natexlab{a}}){Li}, {Chornock}, {Leaman},
  {Filippenko}, {Poznanski}, {Wang}, {Ganeshalingam}, \&
  {Mannucci}}]{2011MNRAS.tmp..317L}
{Li}, W., {Chornock}, R., {Leaman}, J., {et~al.} 2011{\natexlab{a}}, \mnras,
  317

\bibitem[{{Li} {et~al.}(2011{\natexlab{b}}){Li}, {Leaman}, {Chornock},
  {Filippenko}, {Poznanski}, {Ganeshalingam}, {Wang}, {Modjaz}, {Jha}, {Foley},
  \& {Smith}}]{2011MNRAS.tmp..413L}
{Li}, W., {Leaman}, J., {Chornock}, R., {et~al.} 2011{\natexlab{b}}, \mnras,
  413

\bibitem[{{Magnelli} {et~al.}(2009){Magnelli}, {Elbaz}, {Chary}, {Dickinson},
  {Le Borgne}, {Frayer}, \& {Willmer}}]{2009A&A...496...57M}
{Magnelli}, B., {Elbaz}, D., {Chary}, R.~R., {et~al.} 2009, \aap, 496, 57

\bibitem[{{Magnelli} {et~al.}(2011){Magnelli}, {Elbaz}, {Chary}, {Dickinson},
  {Le Borgne}, {Frayer}, \& {Willmer}}]{2011A&A...528A..35M}
{Magnelli}, B., {Elbaz}, D., {Chary}, R.~R., {et~al.} 2011, \aap, 528, A35

\bibitem[{{Maiolino} {et~al.}(2002){Maiolino}, {Vanzi}, {Mannucci}, {Cresci},
  {Ghinassi}, \& {Della Valle}}]{2002A&A...389...84M}
{Maiolino}, R., {Vanzi}, L., {Mannucci}, F., {et~al.} 2002, \aap, 389, 84

\bibitem[{{Mannucci} {et~al.}(2006){Mannucci}, {Della Valle}, \&
  {Panagia}}]{2006MNRAS.370..773M}
{Mannucci}, F., {Della Valle}, M., \& {Panagia}, N. 2006, \mnras, 370, 773

\bibitem[{{Mannucci} {et~al.}(2007){Mannucci}, {Della Valle}, \&
  {Panagia}}]{2007MNRAS.377.1229M}
{Mannucci}, F., {Della Valle}, M., \& {Panagia}, N. 2007, \mnras, 377, 1229

\bibitem[{{Mannucci} {et~al.}(2003){Mannucci}, {Maiolino}, {Cresci}, {Della
  Valle}, {Vanzi}, {Ghinassi}, {Ivanov}, {Nagar}, \&
  {Alonso-Herrero}}]{2003A&A...401..519M}
{Mannucci}, F., {Maiolino}, R., {Cresci}, G., {et~al.} 2003, \aap, 401, 519

\bibitem[{{Maoz} {et~al.}(2011){Maoz}, {Mannucci}, {Li}, {Filippenko}, {Della
  Valle}, \& {Panagia}}]{2011MNRAS.tmp..307M}
{Maoz}, D., {Mannucci}, F., {Li}, W., {et~al.} 2011, \mnras, 307

\bibitem[{{Marcillac} {et~al.}(2006){Marcillac}, {Elbaz}, {Charlot}, {Liang},
  {Hammer}, {Flores}, {Cesarsky}, \& {Pasquali}}]{2006A&A...458..369M}
{Marcillac}, D., {Elbaz}, D., {Charlot}, S., {et~al.} 2006, \aap, 458, 369

\bibitem[{{Mattila} {et~al.}(2012){Mattila}, {Dahlen}, {Efstathiou}, {Kankare},
  {Melinder}, {Alonso-Herrero}, {Perez-Torres}, {Ryder}, {Vaisanen}, \&
  {Ostlin}}]{2012arXiv1206.1314M}
{Mattila}, S., {Dahlen}, T., {Efstathiou}, A., {et~al.} 2012, \apj, accepted,
  [arXiv:1206.1314]

\bibitem[{{Mattila} \& {Meikle}(2001)}]{2001MNRAS.324..325M}
{Mattila}, S. \& {Meikle}, W.~P.~S. 2001, \mnras, 324, 325

\bibitem[{{Mattila} {et~al.}(2004){Mattila}, {Meikle}, \&
  {Greimel}}]{2004NewAR..48..595M}
{Mattila}, S., {Meikle}, W.~P.~S., \& {Greimel}, R. 2004, New Astronomy Review,
  48, 595

\bibitem[{{Mattila} {et~al.}(2007){Mattila}, {V{\"a}is{\"a}nen}, {Farrah},
  {Efstathiou}, {Meikle}, {Dahlen}, {Fransson}, {Lira}, {Lundqvist},
  {{\"O}stlin}, {Ryder}, \& {Sollerman}}]{2007ApJ...659L...9M}
{Mattila}, S., {V{\"a}is{\"a}nen}, P., {Farrah}, D., {et~al.} 2007, \apjl, 659,
  L9

\bibitem[{{Melinder} {et~al.}(2011){Melinder}, {Dahlen}, {Menc{\'{\i}}a
  Trinchant}, {{\"O}stlin}, {Mattila}, {Sollerman}, {Fransson}, {Hayes}, \&
  {Nasoudi-Shoar}}]{2011A&A...532A..29M}
{Melinder}, J., {Dahlen}, T., {Menc{\'{\i}}a Trinchant}, L., {et~al.} 2011,
  \aap, 532, A29

\bibitem[{{Melinder} {et~al.}(2008){Melinder}, {Mattila}, {{\"O}stlin},
  {Menc{\'{\i}}a Trinchant}, \& {Fransson}}]{2008A&A...490..419M}
{Melinder}, J., {Mattila}, S., {{\"O}stlin}, G., {Menc{\'{\i}}a Trinchant}, L.,
  \& {Fransson}, C. 2008, \aap, 490, 419

\bibitem[{{Menc\'ia Trinchant} {et~al.}(2012){Menc\'ia Trinchant}, {Melinder},
  {Dahlen}, {\"Ostlin}, {Fransson}, {Nasoudi-Shoar}, {Hayes}, \&
  {Mattila}}]{laia}
{Menc\'ia Trinchant}, L., {Melinder}, J., {Dahlen}, T., {et~al.} 2012, \aap,
  submitted

\bibitem[{{Miknaitis} {et~al.}(2007){Miknaitis}, {Pignata}, {Rest},
  {Wood-Vasey}, {Blondin}, {Challis}, {Smith}, {Stubbs}, {Suntzeff}, {Foley},
  {Matheson}, {Tonry}, {Aguilera}, {Blackman}, {Becker}, {Clocchiatti},
  {Covarrubias}, {Davis}, {Filippenko}, {Garg}, {Garnavich}, {Hicken}, {Jha},
  {Krisciunas}, {Kirshner}, {Leibundgut}, {Li}, {Miceli}, {Narayan}, {Prieto},
  {Riess}, {Salvo}, {Schmidt}, {Sollerman}, {Spyromilio}, \&
  {Zenteno}}]{2007ApJ...666..674M}
{Miknaitis}, G., {Pignata}, G., {Rest}, A., {et~al.} 2007, \apj, 666, 674

\bibitem[{{Neill} {et~al.}(2006){Neill}, {Sullivan}, {Balam}, {Pritchet},
  {Howell}, {Perrett}, {Astier}, {Aubourg}, {Basa}, {Carlberg}, {Conley},
  {Fabbro}, {Fouchez}, {Guy}, {Hook}, {Pain}, {Palanque-Delabrouille},
  {Regnault}, {Rich}, {Taillet}, {Aldering}, {Antilogus}, {Arsenijevic},
  {Balland}, {Baumont}, {Bronder}, {Ellis}, {Filiol}, {Gon{\c c}alves},
  {Hardin}, {Kowalski}, {Lidman}, {Lusset}, {Mouchet}, {Mourao}, {Perlmutter},
  {Ripoche}, {Schlegel}, \& {Tao}}]{2006AJ....132.1126N}
{Neill}, J.~D., {Sullivan}, M., {Balam}, D., {et~al.} 2006, \aj, 132, 1126

\bibitem[{{Nomoto}(1984)}]{1984ApJ...277..791N}
{Nomoto}, K. 1984, \apj, 277, 791

\bibitem[{{Nomoto} {et~al.}(2009){Nomoto}, {Kamiya}, {Nakasato}, {Hachisu}, \&
  {Kato}}]{2009AIPC.1111..267N}
{Nomoto}, K., {Kamiya}, Y., {Nakasato}, N., {Hachisu}, I., \& {Kato}, M. 2009,
  in AIP Conf. Proc., Vol. 1111, 267--276

\bibitem[{Nugent(2007)}]{2007NugentMISC}
Nugent, P. 2007, Peter Nugent's Spectral Templates,
  \url{http://supernova.lbl.gov/~nugent/nugent_templates.html}

\bibitem[{{Pain} {et~al.}(2002){Pain}, {Fabbro}, {Sullivan}, {Ellis},
  {Aldering}, {Astier}, {Deustua}, {Fruchter}, {Goldhaber}, {Goobar}, {Groom},
  {Hardin}, {Hook}, {Howell}, {Irwin}, {Kim}, {Kim}, {Knop}, {Lee}, {Lidman},
  {McMahon}, {Nugent}, {Panagia}, {Pennypacker}, {Perlmutter}, {Ruiz-Lapuente},
  {Schahmaneche}, {Schaefer}, \& {Walton}}]{2002ApJ...577..120P}
{Pain}, R., {Fabbro}, S., {Sullivan}, M., {et~al.} 2002, \apj, 577, 120

\bibitem[{{Parra} {et~al.}(2007){Parra}, {Conway}, {Diamond}, {Thrall},
  {Lonsdale}, {Lonsdale}, \& {Smith}}]{2007ApJ...659..314P}
{Parra}, R., {Conway}, J.~E., {Diamond}, P.~J., {et~al.} 2007, \apj, 659, 314

\bibitem[{{Pereira-Santaella} {et~al.}(2010){Pereira-Santaella},
  {Alonso-Herrero}, {Rieke}, {Colina}, {D{\'{\i}}az-Santos}, {Smith},
  {P{\'e}rez-Gonz{\'a}lez}, \& {Engelbracht}}]{2010ApJS..188..447P}
{Pereira-Santaella}, M., {Alonso-Herrero}, A., {Rieke}, G.~H., {et~al.} 2010,
  \apjs, 188, 447

\bibitem[{{P{\'e}rez-Torres} {et~al.}(2010){P{\'e}rez-Torres}, {Alberdi},
  {Romero-Ca{\~n}izales}, \& {Bondi}}]{2010A&A...519L...5P}
{P{\'e}rez-Torres}, M.~A., {Alberdi}, A., {Romero-Ca{\~n}izales}, C., \&
  {Bondi}, M. 2010, \aap, 519, L5

\bibitem[{{Perlmutter} {et~al.}(1997){Perlmutter}, {Gabi}, {Goldhaber},
  {Goobar}, {Groom}, {Hook}, {Kim}, {Kim}, {Lee}, {Pain}, {Pennypacker},
  {Small}, {Ellis}, {McMahon}, {Boyle}, {Bunclark}, {Carter}, {Irwin},
  {Glazebrook}, {Newberg}, {Filippenko}, {Matheson}, {Dopita}, {Couch}, \& {The
  Supernova Cosmology Project}}]{1997ApJ...483..565P}
{Perlmutter}, S., {Gabi}, S., {Goldhaber}, G., {et~al.} 1997, \apj, 483, 565

\bibitem[{{Phillips}(1993)}]{1993ApJ...413L.105P}
{Phillips}, M.~M. 1993, \apjl, 413, L105

\bibitem[{{Poznanski} {et~al.}(2007){Poznanski}, {Maoz}, \&
  {Gal-Yam}}]{2007AJ....134.1285P}
{Poznanski}, D., {Maoz}, D., \& {Gal-Yam}, A. 2007, \aj, 134, 1285

\bibitem[{{Pozzo} {et~al.}(2006){Pozzo}, {Meikle}, {Rayner}, {Joseph},
  {Filippenko}, {Foley}, {Li}, {Mattila}, \& {Sollerman}}]{2006MNRAS.368.1169P}
{Pozzo}, M., {Meikle}, W.~P.~S., {Rayner}, J.~T., {et~al.} 2006, \mnras, 368,
  1169

\bibitem[{{Richardson} {et~al.}(2006){Richardson}, {Branch}, \&
  {Baron}}]{2006AJ....131.2233R}
{Richardson}, D., {Branch}, D., \& {Baron}, E. 2006, \aj, 131, 2233

\bibitem[{{Richardson} {et~al.}(2002){Richardson}, {Branch}, {Casebeer},
  {Millard}, {Thomas}, \& {Baron}}]{2002AJ....123..745R}
{Richardson}, D., {Branch}, D., {Casebeer}, D., {et~al.} 2002, \aj, 123, 745

\bibitem[{{Riello} \& {Patat}(2005)}]{2005MNRAS.362..671R}
{Riello}, M. \& {Patat}, F. 2005, \mnras, 362, 671

\bibitem[{{Riess} {et~al.}(2007){Riess}, {Strolger}, {Casertano}, {Ferguson},
  {Mobasher}, {Gold}, {Challis}, {Filippenko}, {Jha}, {Li}, {Tonry}, {Foley},
  {Kirshner}, {Dickinson}, {MacDonald}, {Eisenstein}, {Livio}, {Younger}, {Xu},
  {Dahl{\'e}n}, \& {Stern}}]{2007ApJ...659...98R}
{Riess}, A.~G., {Strolger}, L., {Casertano}, S., {et~al.} 2007, \apj, 659, 98

\bibitem[{{Rodney} \& {Tonry}(2009)}]{2009ApJ...707.1064R}
{Rodney}, S.~A. \& {Tonry}, J.~L. 2009, \apj, 707, 1064

\bibitem[{{Rodney} \& {Tonry}(2010)}]{2010ApJ...723...47R}
{Rodney}, S.~A. \& {Tonry}, J.~L. 2010, \apj, 723, 47

\bibitem[{{Romero-Ca{\~n}izales} {et~al.}(2011){Romero-Ca{\~n}izales},
  {Mattila}, {Alberdi}, {P{\'e}rez-Torres}, {Kankare}, \&
  {Ryder}}]{2011MNRAS.415.2688R}
{Romero-Ca{\~n}izales}, C., {Mattila}, S., {Alberdi}, A., {et~al.} 2011,
  \mnras, 415, 2688

\bibitem[{{Ruiter} {et~al.}(2011){Ruiter}, {Belczynski}, {Sim}, {Hillebrandt},
  {Fryer}, {Fink}, \& {Kromer}}]{2011MNRAS.417..408R}
{Ruiter}, A.~J., {Belczynski}, K., {Sim}, S.~A., {et~al.} 2011, \mnras, 417,
  408

\bibitem[{{Salpeter}(1955)}]{1955ApJ...121..161S}
{Salpeter}, E.~E. 1955, \apj, 121, 161

\bibitem[{{Scannapieco} \& {Bildsten}(2005)}]{2005ApJ...629L..85S}
{Scannapieco}, E. \& {Bildsten}, L. 2005, \apjl, 629, L85

\bibitem[{{Schmidt} {et~al.}(1998){Schmidt}, {Suntzeff}, {Phillips},
  {Schommer}, {Clocchiatti}, {Kirshner}, {Garnavich}, {Challis}, {Leibundgut},
  {Spyromilio}, {Riess}, {Filippenko}, {Hamuy}, {Smith}, {Hogan}, {Stubbs},
  {Diercks}, {Reiss}, {Gilliland}, {Tonry}, {Maza}, {Dressler}, {Walsh}, \&
  {Ciardullo}}]{1998ApJ...507...46S}
{Schmidt}, B.~P., {Suntzeff}, N.~B., {Phillips}, M.~M., {et~al.} 1998, \apj,
  507, 46

\bibitem[{{Sharon} {et~al.}(2010){Sharon}, {Gal-Yam}, {Maoz}, {Filippenko},
  {Foley}, {Silverman}, {Ebeling}, {Ma}, {Ofek}, {Kneib}, {Donahue}, {Ellis},
  {Freedman}, {Kirshner}, {Mulchaey}, {Sarajedini}, \&
  {Voit}}]{2010ApJ...718..876S}
{Sharon}, K., {Gal-Yam}, A., {Maoz}, D., {et~al.} 2010, \apj, 718, 876

\bibitem[{{Shim} {et~al.}(2009){Shim}, {Colbert}, {Teplitz}, {Henry}, {Malkan},
  {McCarthy}, \& {Yan}}]{2009ApJ...696..785S}
{Shim}, H., {Colbert}, J., {Teplitz}, H., {et~al.} 2009, \apj, 696, 785

\bibitem[{{Smartt}(2009)}]{2009ARA&A..47...63S}
{Smartt}, S.~J. 2009, \araa, 47, 63

\bibitem[{{Smartt} {et~al.}(2009){Smartt}, {Eldridge}, {Crockett}, \&
  {Maund}}]{2009MNRAS.395.1409S}
{Smartt}, S.~J., {Eldridge}, J.~J., {Crockett}, R.~M., \& {Maund}, J.~R. 2009,
  \mnras, 395, 1409

\bibitem[{{Strolger} {et~al.}(2010){Strolger}, {Dahlen}, \&
  {Riess}}]{2010ApJ...713...32S}
{Strolger}, L., {Dahlen}, T., \& {Riess}, A.~G. 2010, \apj, 713, 32

\bibitem[{{Strolger} {et~al.}(2004){Strolger}, {Riess}, {Dahlen}, {Livio},
  {Panagia}, {Challis}, {Tonry}, {Filippenko}, {Chornock}, {Ferguson},
  {Koekemoer}, {Mobasher}, {Dickinson}, {Giavalisco}, {Casertano}, {Hook},
  {Blondin}, {Leibundgut}, {Nonino}, {Rosati}, {Spinrad}, {Steidel}, {Stern},
  {Garnavich}, {Matheson}, {Grogin}, {Hornschemeier}, {Kretchmer}, {Laidler},
  {Lee}, {Lucas}, {de Mello}, {Moustakas}, {Ravindranath}, {Richardson}, \&
  {Taylor}}]{2004ApJ...613..200S}
{Strolger}, L., {Riess}, A.~G., {Dahlen}, T., {et~al.} 2004, \apj, 613, 200

\bibitem[{{Sullivan} {et~al.}(2006){Sullivan}, {Le Borgne}, {Pritchet},
  {Hodsman}, {Neill}, {Howell}, {Carlberg}, {Astier}, {Aubourg}, {Balam},
  {Basa}, {Conley}, {Fabbro}, {Fouchez}, {Guy}, {Hook}, {Pain},
  {Palanque-Delabrouille}, {Perrett}, {Regnault}, {Rich}, {Taillet}, {Baumont},
  {Bronder}, {Ellis}, {Filiol}, {Lusset}, {Perlmutter}, {Ripoche}, \&
  {Tao}}]{2006ApJ...648..868S}
{Sullivan}, M., {Le Borgne}, D., {Pritchet}, C.~J., {et~al.} 2006, \apj, 648,
  868

\bibitem[{{Totani} {et~al.}(2008){Totani}, {Morokuma}, {Oda}, {Doi}, \&
  {Yasuda}}]{2008PASJ...60.1327T}
{Totani}, T., {Morokuma}, T., {Oda}, T., {Doi}, M., \& {Yasuda}, N. 2008,
  \pasj, 60, 1327

\bibitem[{{Tremonti} {et~al.}(2007){Tremonti}, {Moustakas}, \&
  {Diamond-Stanic}}]{2007ApJ...663L..77T}
{Tremonti}, C.~A., {Moustakas}, J., \& {Diamond-Stanic}, A.~M. 2007, \apjl,
  663, L77

\bibitem[{{Trenti} \& {Stiavelli}(2008)}]{2008ApJ...676..767T}
{Trenti}, M. \& {Stiavelli}, M. 2008, \apj, 676, 767

\bibitem[{{Tsujimoto} {et~al.}(1997){Tsujimoto}, {Yoshii}, {Nomoto},
  {Matteucci}, {Thielemann}, \& {Hashimoto}}]{1997ApJ...483..228T}
{Tsujimoto}, T., {Yoshii}, Y., {Nomoto}, K., {et~al.} 1997, \apj, 483, 228

\bibitem[{{Zwicky}(1938)}]{1938ApJ....88..529Z}
{Zwicky}, F. 1938, \apj, 88, 529

\end{thebibliography}
                                                   
\end{document}